\documentclass{pasa}%
\usepackage{aas_macros,amssymb}
\usepackage[authoryear]{natbib}
\bibpunct{(}{)}{;}{a}{}{,}
\title[Observable Imprints of Small-Scale Structure on Dark Matter Haloes]
	{Seeking Observable Imprints of Small-Scale Structure on the 
          Properties of Dark Matter Haloes}
\author[C. Power]{C. Power\thanks{chris.power@icrar.org}
\affil{International Centre for Radio Astronomy Research, University of 
  Western Australia, 35 Stirling Highway, Crawley, Western Australia 6009, 
  Australia}}
\jid{PASA}
\doi{10.1017/pas.\the\year.xxx}
\jyear{\the\year}

\newcommand{\gae}{\lower 2pt \hbox{$\, \buildrel {\scriptstyle >}\over {\scriptstyle
\sim}\,$}}
\newcommand{\lae}{\lower 2pt \hbox{$\, \buildrel {\scriptstyle <}\over {\scriptstyle
\sim}\,$}}

\begin{document}

\begin{abstract}

The characteristic prediction of the Cold Dark Matter (CDM) model of 
cosmological structure formation is that the Universe should contain 
a wealth of small-scale structure -- low-mass dark matter haloes and
subhaloes. However, galaxy formation is inefficient in their shallow 
potential wells and so we expect these low-mass haloes and subhaloes
to be dark. Can we tell the difference between a Universe in which 
these low-mass haloes are present but dark and one in which they never 
formed, thereby providing a robust test of the Cold Dark Matter model? 
We address this question using cosmological $N$-body simulations to 
examine how properties of low-mass haloes that are potentially accessible 
to observation, such as their spatial clustering, rate of accretions and 
mergers onto massive galaxies and the angular momentum content of
massive galaxies, differ between a fiducial $\Lambda$CDM model and 
dark matter models in which low-mass halo formation is suppressed. Adopting
an effective cut-off mass scale $M_{\rm cut}$ below which small-scale power 
is suppressed in the initial conditions, we study dark matter models in which 
$M_{\rm cut}$ varies between $5 \times 10^9\,h^{-1} {\rm M_{\odot}}$ and 
$10^{11}\, h^{-1} {\rm M_{\odot}}$, equivalent to the host haloes of dwarf 
and low mass galaxies. Our results show that both the clustering strength 
of low-mass haloes around galaxy-mass primaries and the rate at which they
merge with these primaries is sensitive to the assumed value of $M_{\rm cut}$; 
in contrast, suppressing low-mass halo formation has little influence on the 
angular momentum content of galaxy-mass haloes -- it is the quiescence or 
violence of a halo's assembly history that has a more marked effect. However, 
we expect that measuring the effect on spatial clustering or the merger rate 
is likely to be observationally difficult for realistic values of $M_{\rm cut}$, 
and so isolating the effect of this small-scale structure would appear to be 
remarkably difficult to detect, at least in the present day Universe.
\end{abstract}

\begin{keywords}
  methods: $N$-body simulations -- galaxies: formation -- galaxies:
  haloes -- cosmology: theory -- dark matter -- large-scale structure of Universe

\end{keywords}

\maketitle

\section{Introduction}
\label{sec:introduction}


One of the key questions facing fundamental physics and cosmology at the 
turn of the $21^{\rm st}$ century concerns the nature of the dark matter. 
Approximately $80\%$ of the matter content of the Universe appears to be 
in the form of exotic, non-baryonic dark matter \citep[cf.][]{planck.2013}
whose clustering is believed to play a crucial role in the formation and 
subsequent evolution of galaxies \citep[e.g.][]{white78,white91}. 
This non-baryonic dark matter is widely assumed to be cold -- that is, dark 
matter particles were non-relativistic at the time of decoupling -- and 
collisionless, and these properties of Cold Dark Matter (hereafter CDM) 
lead to a number of fundamental consequences. The first of these is that dark 
matter haloes have central density cusps \citep[cf.][]{tremaine79,moore94}; 
the second is that the halo mass function -- the number of haloes of mass $M$ 
per unit mass per unit comoving volume -- increases with decreasing mass as 
$M^{-\alpha}$ where $\alpha \sim 2.0$ 
\citep[cf. Table 1 of ][]{murray.etal.2013} down to masses that could be as 
small as $\sim 10^{-6} \rm M_{\odot}$ \citep[cf.][]{green04}.

Based on the results of cosmological $N$-body simulations, cuspy haloes and 
an abundance of small-scale structure -- low-mass haloes and subhaloes -- are 
now well established as robust predictions of the CDM model 
\citep[e.g.][]{springel.2008}. If we consider a Universe in which low-mass 
halo and subhalo formation is suppressed, as in Warm Dark Matter 
(hereafter WDM) models, we find that haloes can form cores 
\citep[albeit small ones; cf.][]{villaescusa-navarro.dalal.2011} but they
are likely to remain cuspy for plausible WDM particle masses 
($m_{\rm WDM} \gtrsim 0.5 - 2\,\rm keV$), even when free streaming is accounted 
for \citep[e.g.][]{colin08,maccio.etal.2012a}. In contrast, we expect the 
abundance and clustering of low-mass haloes to be suppressed in WDM models
\citep[e.g.][]{dunstan.etal.2011,smith.markovic.2011,schneider.2013,benson.2013,pacucci.2013}, and so it can be argued 
that it is the abundance of small-scale structure, rather than
central density cusps, that is the defining characteristic of the CDM 
model\footnote{There is a caveat here -- we have assumed implicitly that dark 
matter is collisionless. Recent papers have revived the possibility
that dark matter is self-interacting 
\citep[e.g.][]{loeb.weiner.2011,vogelsberger.etal.2012,peter.etal.2013,rocha.etal.2013}, an idea that was 
first explored systematically using simulations in the early 2000s 
\citep[see, for example][]{yoshida00,dave01,colin02}. This recent work has 
shown that it is possible to form small cores ($r_{\rm core} \simeq 1\,\rm kpc$) 
in Milky Way mass galaxies \citep{loeb.weiner.2011} while leaving the abundance
and radial distribution of dark matter subhaloes unchanged from the prediction 
of the CDM model \citep[e.g.][]{vogelsberger.etal.2012,rocha.etal.2013}.}. 

However, we expect few of these low-mass haloes to host galaxies because 
galaxy formation will be inefficient in their shallow potential wells
\citep[e.g.][]{dekel86,efstathiou92,thoul96,benson02a}. For example, 
supernovae \citep[e.g.][]{dekel86} and photo-ionizing sources 
\citep[e.g.][]{benson02a,cantalupo.2010} can quench galaxy formation in 
low-mass haloes, while the likelihood that a low-mass halo hosts a satellite 
galaxy appears to be stochastic 
\citep[cf.][]{boylan-kolchin.etal.2012,garrison-kimmel.etal.2013}, suggesting 
that the process is highly sensitive to details of a galaxy's assembly 
history (i.e. environment, gas accretion and star formation history, etc...). 
This raises the question, if low-mass haloes and subhaloes remain dark because 
galaxy formation is inefficient on these mass scales, how can we tell the 
difference between a Universe in which small-scale structure is present but 
dark and one in which its formation is suppressed, as in WDM models?

In this paper we use the results of cosmological $N$-body simulations to 
address this question, comparing systematically dark matter halo properties
in a fiducial $\Lambda$CDM cosmology and in $\Lambda$WDM-like dark matter 
models, in which low-mass halo formation is suppressed by truncating the 
$\Lambda$CDM power spectrum on small scales. Numerous studies have 
investigated the halo mass function in WDM models \citep[recent examples include][]{schneider.etal.2011,schneider.2013,pacucci.2013,benson.2013} and 
associated issues arising from discreteness effects in such simulations 
\citep[e.g.][]{wang07,schneider.2013,angulo.2013,hahn.2013}, but we note that
direct measurement of the halo mass function observationally is fraught with
difficulty \citep[cf.][]{murray.etal.2013b}. For this reason we focus on three 
measures of the halo population that are potentially accessible to observation 
-- (i) the spatial clustering of low-mass haloes around galaxy- and group-mass 
haloes ($10^{11} h^{-1}{\rm M}_{\odot}\lae\,M_{\rm vir}\lae10^{13} h^{-1}{\rm M}_{\odot}$); (ii) 
the rate at which these haloes assemble their mass and at which they 
experience mergers; and (iii) their angular momentum content. Although we 
analyse properties of the halo population, we reason that they provide a 
baseline for trends that we observe in the galaxy population.

We choose the cut-off mass $M_{\rm cut}$, the mass scale below which halo 
formation is suppressed, to vary between $5 \times 10^9 h^{-1} {\rm M_{\odot}} 
\lae M_{\rm cut} \lae 10^{11} h^{-1} {\rm M_{\odot}}$. These values of 
$M_{\rm cut}$ are unrealistic in the sense that they are too large to be 
consistent with observational constraints (see, for example, the review of 
\citealt{primack09}, assuming corresponding filtering masses from 
\citealt{bode01}) but they allow us to experiment with the consequences of 
progressively more aggressive truncations of the initial power spectrum on 
the properties of haloes with $M\!\gtrsim\!10^{11} \rm M_{\odot}$.

The structure of this paper is as follows. In \S\ref{sec:simulations} we 
describe our simulations, detailing how we set them up and summarising our 
approach to constructing merger trees and halo sample selection. In 
\S\ref{sec:res_mf_clus} we focus on the spatial clustering of haloes and 
demonstrate that the clustering strength of low-mass haloes is suppressed 
relative to the $\Lambda$CDM model in our truncated models. We show how this suppression in clustering impacts on the number and frequency of minor mergers 
(\S\ref{sec:res_accretion}) and we explore measures 
of halo angular momentum and spin (\S\ref{sec:res_spin_growth}). Finally, in 
\S\ref{sec:conclusions} we summarise our results, assessing their 
implications for developing robust 
astrophysical tests of the nature of the dark matter.

\section{The Simulations}
\label{sec:simulations}

We have run a sequence of cosmological $N$-body simulations that follow
the formation and evolution of dark matter haloes in a box of side
$20 h^{-1} \rm Mpc$ from a starting redshift of $z$=100 to $z$=0.
For each run we assume a flat cosmology with a dark energy term, 
and for convenience we adopt the cosmological parameters of 
\citet{wmap07} -- matter and dark energy density parameters of
$\Omega_{\rm m}=0.24$ and $\Omega_{\Lambda}=0.76$, a dimensionless 
Hubble parameter of $h=0.73$, a normalisation of $\sigma_8=0.74$ 
and a primordial spectral index of $n_{\rm spec}$=0.95. Each simulation volume
contains $256^3$ equal-mass particles, which for the adopted cosmological
parameters gives particle masses of $m_{\rm p}=3.176 \times 10^7 h^{-1} \rm
M_{\odot}$. 

The respective runs differ in the spatial scale below which small-scale 
power in the initial conditions is suppressed. We generate a single
realisation of the $\Lambda$CDM power spectrum appropriate for our choice of
cosmological parameters and in the case of the truncated models we 
introduce a sharp cut-off in the $\Lambda$CDM power spectrum at progressively 
larger spatial scales. This cut-off spatial scale is set by the mass scale 
below which we wish to suppress halo formation. Details about the truncated
models are given in the next section.

All of our simulations were run using the parallel TreePM code {\small GADGET2}
\citep{gadget2} with a constant comoving gravitational softening 
$\epsilon$=$1.5\,h^{-1}\rm kpc$ and individual and adaptive particle 
time-steps. These were assigned according to the criterion 
$\Delta t = \eta \sqrt{\epsilon/a}$, where $a$ is the magnitude of 
a particle's gravitational acceleration and $\eta=0.05$ determines 
the accuracy of the time integration.

\subsection{Truncated Dark Matter Models}

\subsubsection*{Truncating the Initial Power Spectrum}

We are interested in models in which small scale power is suppressed 
at early times. Physically suppression arises because dark
matter free streams, which acts as a damping mechanism to wash out
primordial density perturbations and to introduce a cut-off in the linear 
matter power spectrum. If the dark matter particle is a thermal relic,
the spatial scale at which this cut-off occurs can be calculated 
\citep[cf.][]{bergstrom00}. The free streaming scale $\lambda_{\rm fs}$ 
can be expressed as 
\begin{equation}
  \label{eq:lambda_fs}
  \lambda_{\rm fs} = 0.2\,(\Omega_{\rm dm}\,h^2)^{1/3}\,\left(\frac{m_{\rm dm}}{{\rm keV}}\right)^{-4/3} \,\rm Mpc,
\end{equation}
\noindent where $m_{\rm dm}$ is the dark matter particle mass measured in
$\rm keV$ and $\Omega_{\rm dm}$ is the dark matter density
\citep[cf.][]{boehm05}. 

Provided $\lambda_{\rm fs}$ is small
compared to the spatial scales we are interested in simulating,
the power spectrum will differ little from the $\Lambda$CDM power spectrum 
\citep[which itself may have a cut-off on comoving scales of order 1
  pc; cf.][]{green04}. However, as $\lambda_{\rm fs}$ increases and approaches 
the scale that we wish to resolve, then it becomes necessary to determine how 
the power spectrum changes. The shape of the linear power spectrum for 
collisionless WDM models has been calculated by a number of authors 
\citep[e.g.][]{bbks,bode01}, and it can be recovered from the CDM power 
spectrum by introducing an exponential cut-off at small scales. The 
larger $\lambda_{\rm fs}$, the larger the mass scale $\rm M_{\rm fs}$ 
below which structure formation is suppressed and the smaller the 
wave-number $k$ at which the WDM and CDM power spectra differ, although 
the relationship between $\lambda_{\rm fs}$ and $\rm M_{\rm fs}$ is 
sensitive to the precise nature of the WDM particle.

We do not wish to make any assumptions about the precise nature 
of the dark matter other than that it is collisionless and that low-mass 
halo formation is suppressed, and so we follow \citet{moore99a} and 
truncate sharply the power spectrum at $k_{\rm cut}$, suppressing power at 
wave-numbers $k \geqslant k_{\rm cut}$. We choose $k_{\rm cut}$ by identifying 
a mass scale 
$M_{\rm cut}$ and estimating the comoving length scale $R_{\rm cut}$,

\begin{equation}
  \label{eq:rcut}
  R_{\rm cut} = \left(\frac{3\,M_{\rm cut}}{4\pi}\frac{1}{\overline{\rho}}\right)^{1/3}
\end{equation}

\noindent where $\overline{\rho}$ is the mean density of the Universe. 

\subsubsection*{Modelling Free Streaming}

Similarly we choose not to include the effect of free streaming in 
our initial conditions -- partly because we wish to avoid assumptions 
about the precise nature of the dark matter, and partly for pragamtic 
reasons, which we now explain. In practice free streaming is mimicked by 
assigning a random velocity component (typically drawn from a 
Fermi-Dirac distribution) to particles in addition to their velocities 
predicted by linear theory 
\citep[cf.][]{klypin93,colin08,maccio.etal.2012a}. However, capturing 
this effect correctly in a $N$-body simulation is difficult -- it can 
lead to an unphysical excess of small-scale power in the initial 
conditions if the simulation is started too early (see Figure 1 of 
\citealt{colin08} for a nice illustration of this problem).

Precisely how early is too early has yet to be properly quantified, but 
it will depend explicitly on dark matter particle mass -- the lower the 
mass, the longer the free streaming scale, and the larger the random 
velocity component required. If this exceeds the typical velocity 
predicted by linear theory, a population of spurious haloes forms 
\citep[e.g.][]{klypin93} that can exceed in number by factors of $\sim$ 
10 the population that forms when no random velocity component is 
included, as studied by \citet[][]{wang07}. This is unlikely to be a 
problem for studying the mass profiles of haloes -- for example, 
\citet{colin08} started their simulations at reasonably late times 
because the random velocity component damps away with decreasing 
redshift while the velocities predicted by linear theory increase -- but 
it is not clear how much of a problem it will be for our study, in which 
we study quantities that depend on spatial correlations between haloes.
For this reason we do not include the effect of free streaming, deferring 
this to a forthcoming study on discreteness effects in WDM-like simulations
(Power et al., in preparation).

\subsubsection*{Generation of Initial Conditions}


We follow the standard procedure \citep[e.g.][]{power03} of generating 
a statistical
realisation of the high redshift density field using the appropriate
linear theory power spectrum, from which initial displacements and 
velocities are computed and imposed on a suitable uniform particle 
load; for this study we adopt an initial glass distribution
\citep[cf.][]{white94}. We use the Boltzmann code {\small
  CMBFAST} \citep[][]{cmbfast} to generate the CDM transfer function
for our choice of cosmological parameters. This is convolved with the
primordial power spectrum ($P(k)\propto k^n$, where $n$ is the primordial
spectral index) to obtain the appropriate $\Lambda$CDM power 
spectrum $P(k)$. To obtain a truncated model, we chop $P(k)$ sharply 
at $k_{\rm cut} = 2\pi/R_{\rm cut}$ (where $R_{\rm cut}$ is given by 
equation~\ref{eq:rcut}) and thereby suppress power on scales 
$k \gtrsim k_{\rm cut}$.

\begin{table}
\begin{center}
  \caption{\textbf{Truncated Models : Simulation Details}}
\vspace*{0.3 cm}

\begin{tabular}{lccc}
\hline
Model & $M_{\rm cut}$ & $R_{\rm cut}$ & $k_{\rm cut}$ \\
      & $10^{10} h^{-1} \rm M_{\odot}$ & $h^{-1} \rm Mpc$ & $h {\rm Mpc}^{-1}$ \\

\hline

A &  0.5 & 0.26 & 24.01 \\
B &  1.0 & 0.33 & 19.06 \\
C &  5.0 & 0.56 & 11.15 \\
D & 10.0 & 0.71 &  8.85 \\

\hline
\end{tabular}
\label{tab:details}
\end{center}
\end{table}

We consider five cases -- a fiducial $\Lambda$CDM model and truncated
models in which small scale power is suppressed at masses below 
$M_{\rm cut}=5\times 10^9$, $10^{10}$, $5 \times 10^{10}$ and 
$10^{11} h^{-1} M_{\odot}$ respectively. Note that the cut-off wave-number 
$k_{\rm cut}$ is always less than the Nyquist frequency of the simulation, 
$k_{\rm Ny} \simeq 40 h\, \rm Mpc^{-1}$. Values for the cut-off masses and
wave-numbers are given in Table~\ref{tab:details}. 

\subsection{Halo Identification \& Merger Trees}

\subsubsection*{Halo Identification} 

Groups are identified using {\small AHF}, 
otherwise known as {\small \textbf{A}MIGA}'s {\small \textbf{H}}alo {\small
  \textbf{F}}inder \citep[cf.][]{ahf}. {\small AHF} locates groups as 
peaks in an adaptively smoothed density field using a hierarchy of grids 
and a refinement criterion that is comparable to the force resolution of 
the simulation. Local potential minima are 
calculated for each of these peaks and the set of particles that are 
gravitationally bound to the peaks are identified as the groups that form 
our halo catalogue. Each halo in the catalogue is then processed, 
producing a range of structural and kinematic information. 

We adopt the standard definition of a halo such that the virial 
mass is 
\begin{equation}
  \label{eq:mvir}
  M_{\rm vir}=4 \pi \rho_{\rm crit} \Delta_{\rm vir} r_{\rm vir}^3/3, 
\end{equation}
\noindent where $\rho_{crit}=3H^2/8\pi G$ is the critical density of the 
Universe and $r_{\rm vir}$ is the virial radius, which defines the radial 
extent of the halo. The virial over-density criterion, $\Delta_{\rm vir}$, 
is a multiple of the critical density, and corresponds to 
the mean over-density at the time of virialisation in the spherical collapse 
model \citep[the simplest analytic model of halo formation; cf. ][]{eke96}.
In an Einstein-de Sitter Universe, $\Delta_{\rm vir} \simeq 178$, while in
the favoured $\lambda$CDM model $\Delta_{\rm vir} \simeq 92$ at $z$=0.

Defined in this way, the virial radius $r_{\rm vir}$ provides a convenient 
albeit approximate boundary for a dark matter halo that can be estimated 
easily from simulation data. However, it is \emph{only} approximate -- 
haloes that form in cosmological simulations are relatively complex
structures. They are generally aspherical \citep[e.g.][]{bailin05} and 
asymmetric \citep[e.g.][]{gao06} with no simple boundary 
\citep[e.g.][]{prada06}, and so defining an appropriate boundary is not 
straightforward. This presents difficulties when calculating, for example, 
a halo's angular momentum and its binding energy \citep[cf.][]{lokas01}. 
Material bound to the halo can lie outside of $r_{\rm vir}$, and this will 
distort the angular momentum and binding energy one measures for the 
\emph{halo} using only material from within $r_{\rm vir}$. This issue has been
touched on by previous authors \citep[e.g.][]{cole96,lokas01,shaw06,power11} in the 
context of identifying when a halo is in virial equilibrium. In a similar vein,
the angular momentum one measures using only material from within $r_{\rm
  vir}$ will be biased. This is an important caveat that we need to bear in
mind when discussing our analysis of halo angular momentum in 
\S~\ref{sec:res_spin_growth}.

\subsubsection*{Halo Merger Trees} 

Halo merger trees are constructed by 
linking halo particles at consecutive output times;

\begin{itemize}
\item For each pair of group catalogues constructed at consecutive
  output times $t_1$ and $t_2>t_1$, the `ancestors' of 'descendant'
  groups are identified. For each descendent identified in the catalogue at 
  the later time $t_2$, we sweep over its associated particles and
  locate every ancestor at the earlier time $t_1$ that contains a subset
  of these particles. A record of all ancestors at $t_1$ that contain
  particles associated with the descendent at $t_2$ is maintained.

\item The ancestor at time $t_1$ that contains in excess of $f_{\rm prog}$
  of these particles and also contains the most bound particle of the
  descendent at $t_2$ is deemed the main progenitor. Typically $f_{\rm
  prog}=0.5$, i.e. the main progenitor contains in excess of half the
  final mass.

\end{itemize}

\noindent Each group is then treated as a node in a tree structure, which 
can be traversed either forwards, allowing one to identify a halo at some
early time and follow it forward through the merging hierarchy, or
backwards, allowing one to identify a halo and all its progenitors at
earlier times. In our analysis we concentrate on the main trunk of
the merger tree, in which we follow the evolution of the main
progenitor of a halo to earlier times.

\subsection{Selecting the Halo Sample}

However, care must be taken when including haloes with masses below 
$\rm M_{\rm cut}$ in any analysis. An unfortunate feature of simulations of 
cosmologies in which small-scale power is suppressed at early times is the 
formation of unphysical low-mass haloes by the fragmentation of filaments, 
driven by the discreteness of the matter distribution. These spurious 
haloes form preferentially in filaments, at regular intervals of order the 
mean interparticle separation of the simulation, akin to ``beads on a 
string'' (see below). The mass scale below which these spurious ``haloes'' form can be 
estimated from the halo mass function as a sharp upturn in the number 
density, and it has been shown to scale as ${\rm M_{\rm lim}} \sim 3.9\, 
m_p^{1/3} {\rm M_{\rm cut}}^{2/3}$, where $m_p$ is the particle mass 
\citep[][]{wang07}.

We wish to identify haloes in the truncated models that have clearly
identifiable counterparts in the fiducial $\Lambda$CDM
simulation. These haloes form the halo sample upon which our analysis
is based. By selecting haloes in this way, we can track the merger
trees of the counterparts and study the merging and accretion histories 
of individual systems, correlating any differences in halo properties
with the details of their mass assembly. We can also avoid including in
our analysis spurious (unphysical) haloes that form below the mass
cut-off in the truncated models \citep[see, for example,][]{wang07}.

To identify counterparts, we adapt our algorithm for linking haloes
across time slices when building merger trees to link haloes between 
runs at a given time.

\begin{itemize}
\item For each pair of group catalogues, we process each group and
  compute ``virial'' quantities, namely the virial mass and radius, and
  the set of particles that belong to each halo.

\item For each halo in the fiducial $\Lambda$CDM model at time $t$, we
  loop over its associated particles and determine how many of these
  particles are present in haloes in the corresponding truncated
  model catalogue.

\item The halo in the truncated model that contains in excess of 
  $f_{\rm count}=75\%$ of these particles is identified as a
  counterpart candidate. However, the candidate halo can be part of a
  much larger structure in the fiducial $\Lambda$CDM model, so we also
  require that the mass of the candidate halo not differ from its CDM
  counterpart by not more than a factor of $25\%$. Haloes that satisfy
  these conditions are identified as counterparts.
\end{itemize}

\section{Spatial Clustering}
\label{sec:res_mf_clus}



As our starting point, we compare and contrast the spatial clustering of 
dark matter haloes in the $\Lambda$CDM and truncated models 
respectively as a function of redshift. We expect differences between 
models to be apparent for haloes with masses $M \sim M_{\rm cut}$ and to 
become more pronounced with increasing redshift, when $M_{\rm cut}$ is a 
larger fraction of the typical collapsing mass $M^{*}$.

\subsubsection*{Visual Impression} 

\begin{figure*}
  \centering
  \includegraphics[width=16cm]{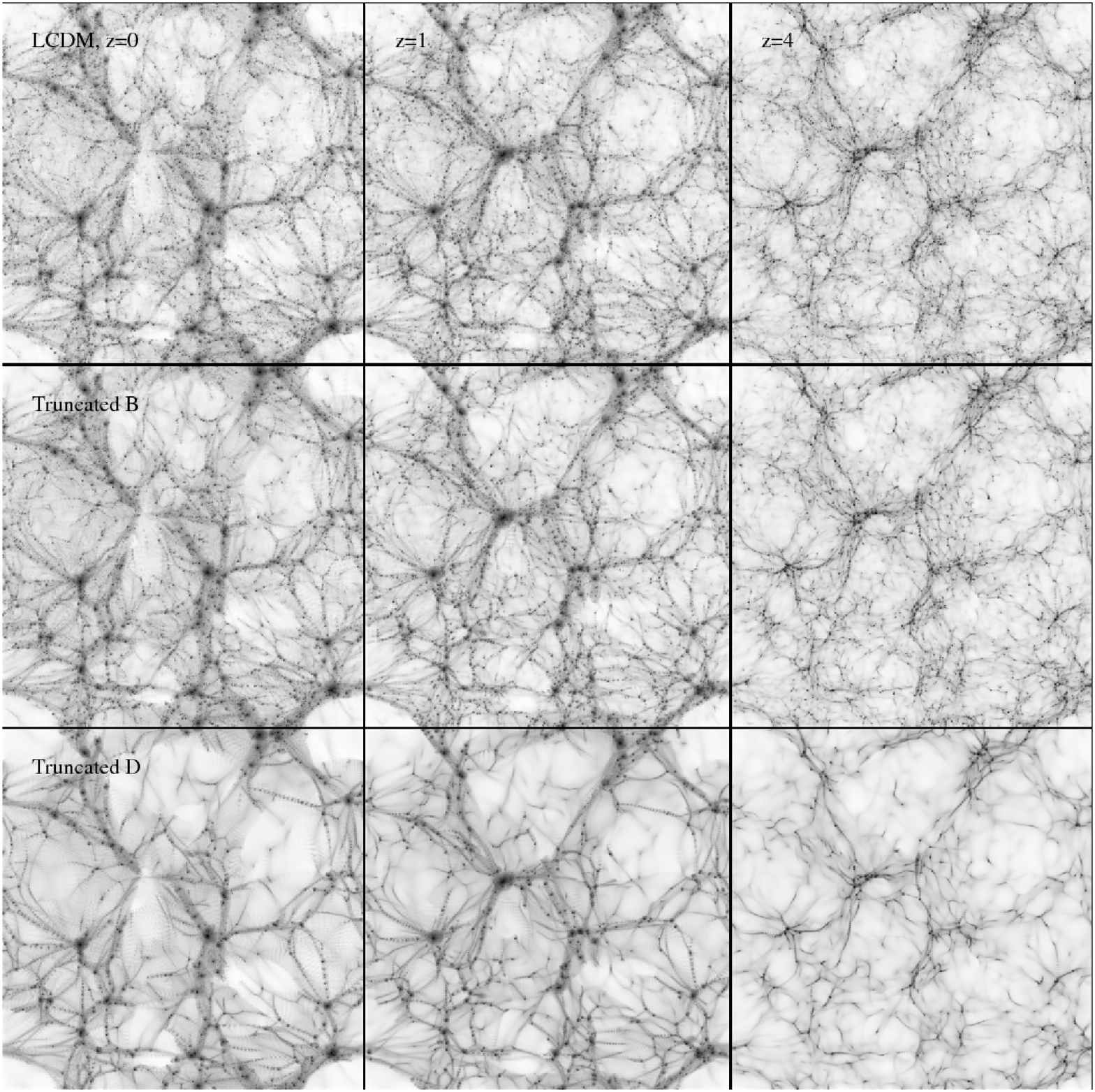}
  \caption{The projected density distribution in $2 h^{-1} \rm Mpc$
  slices taken through the centres of each of the boxes. We have
  smoothed the particle mass using an adaptive Gaussian kernel and
  projected onto a mesh. Each mesh point is weighted according to the
  logarithm of its projected surface density, and so the ``darker'' the
  mesh point, the higher the projected surface density.}
\label{fig:slices}
\end{figure*}

In Figure~\ref{fig:slices} we show the projected 
dark matter distribution in thin slices ($20 \times 20 \times 2 h^{-3} 
\rm Mpc^3$) taken through the $\Lambda$CDM (upper panels), Truncated B 
(middle panels; hereafter TruncB) and D (lower panels; hereafter TruncD) 
at $z$ = 0, 1 and 4 (from left to right). Each slice is centred on the 
geometric centre of the simulation volume 
and the grey-scale is weighted by the logarithm of projected density. 

Figure~\ref{fig:slices} is instructive because it provides a powerful 
visual impression of the effect of suppressing small scale power at 
early times. The filamentary network is largely unaffected and the 
positions of the most massive haloes, which form at the nodes of these 
filaments, are similar in each of the models we have looked at. What is 
striking, however, is the impact on the abundance of low-mass haloes, 
which appear as small dense knots in projection. As $M_{\rm cut}$ 
increases, the projected number density of these low-mass haloes 
decreases markedly as we go from the $\Lambda$CDM run to the TruncD run 
(top and bottom panels respectively). This is evident in the clustering 
around more massive haloes and the absence of low-mass systems in the 
void regions. Furthermore, the contrast between the models becomes 
increasingly noticeable with increasing redshift -- compare $z$=0 and 
$z$=4. Note also the presence of the low-mass haloes distributed along 
filaments in ``beads-on-a-string'' fashion in the truncated models.

\begin{figure}
  \centering
  \includegraphics[width=\columnwidth]{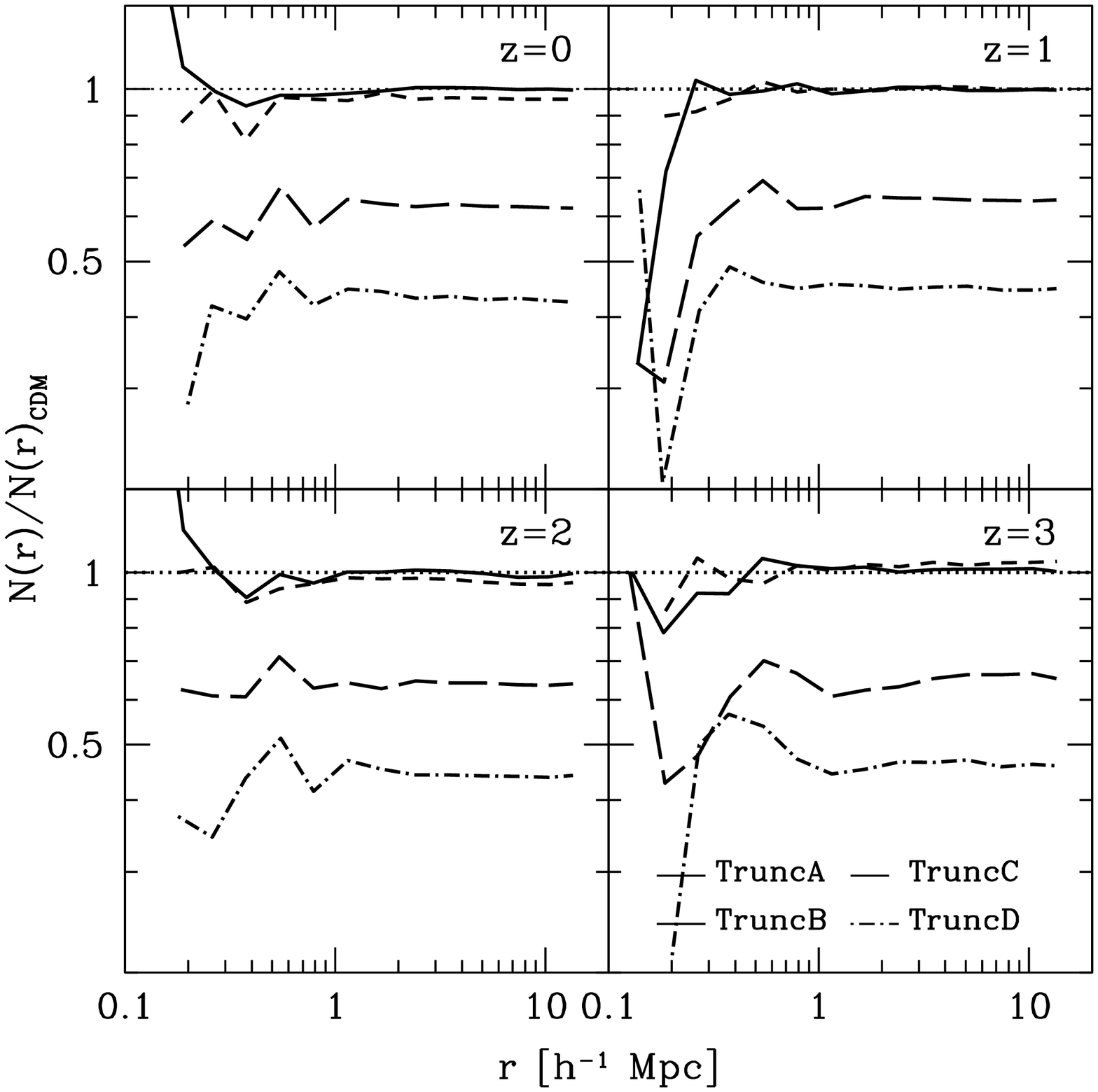}
  \centering
  \includegraphics[width=\columnwidth]{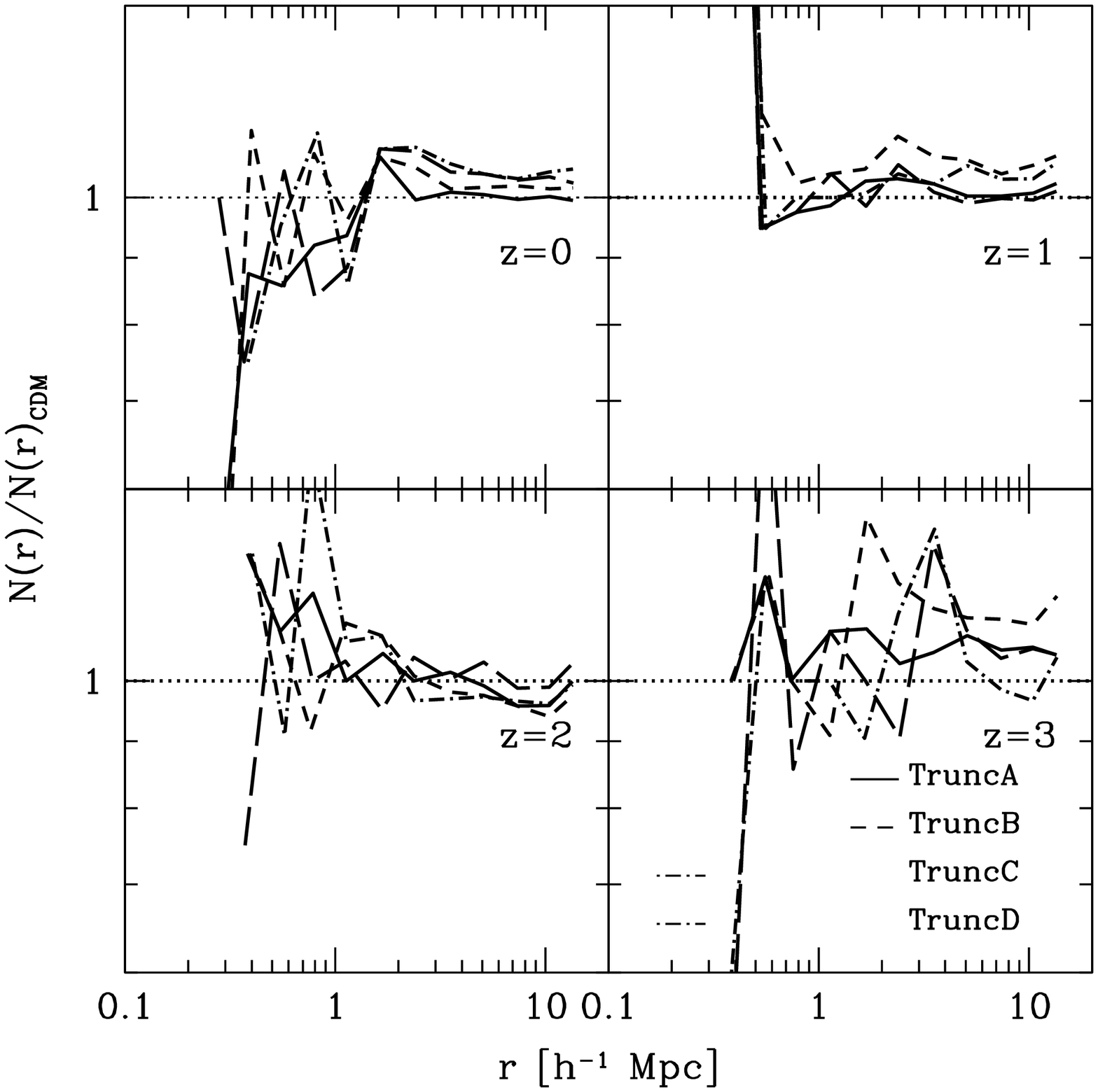}
  \caption{{\bf Evolution of Spatial Clustering with Redshift.} 
    We examine how the clustering strength of haloes with respect to the fiducial
    $\Lambda$CDM model varies across the runs TruncA (solid curves), TruncB (short dashed 
    curves), TruncC (long dashed curves) and TruncD (dotted-dashed curves) at 
    $z$=0, 1, 2 and 3 by plotting the ratio $N(r)/N(r)_{\rm CDM}$ -- the
    number of haloes with comoving halo separation $r$ -- as a function of $r$.
    In the upper panel, we look at the clustering of all secondary haloes with 
    mass $M_{\rm vir} \geqslant 3 \times 10^{9} h^{-1} \rm  M_{\odot}$ around primary
    haloes with mass $M_{\rm vir} \geqslant 10^{11} h^{-1} \rm M_{\odot}$, while in the 
    lower panel we look at the clustering of only massive haloes, for which both 
    the primary and secondary masses are $M_{\rm vir} \geqslant 10^{11} h^{-1} \rm M_{\odot}$.}
  \label{fig:clustering}
\end{figure}

\subsubsection*{Spatial Clustering} 

In Figure~\ref{fig:clustering} we 
investigate how the clustering strength of haloes differs between the 
different dark matter models and as a function of redshift. We quantify 
clustering strength by the correlation function $\xi(r)$, which measures 
the excess probability over random that a pair of haloes $i$ and $j$ 
will be separated by a distance $r=|\vec{r}|=|\vec{r}_i-\vec{r}_j|$. 
$\xi(r)$ is estimated using

\begin{equation}
  \label{eq:xiofr}
  \xi(r)=1+\frac{\overline{DD(r)}}{\overline{RR(r)}},
\end{equation}

\noindent where $\overline{DD(r)}$ is the number of haloes at 
comoving separation $r$ compared to the number in a random distribution 
$\overline{RR(r)}$. Because our focus is on differences, we construct the ratio of 
$N(r)=\overline{DD(r)}=\overline{RR(r)}(\xi(r)-1)$ for each truncated model
to $N(r)_{\rm CDM}$ for the fiducial $\Lambda$CDM run. 

In the upper panel of Figure~\ref{fig:clustering} we consider pairs of 
haloes in which the primary's mass is $M_{\rm vir} \geqslant 10^{11} h^{-1} \rm M_{\odot}$ 
and the secondary's mass is $M_{\rm vir} \geqslant 3 \times 10^{9} h^{-1} \rm 
M_{\odot}$, while in the lower panel we consider pairs of haloes in which
both the primary and secondary masses $M_{\rm vir} \geqslant 10^{11} h^{-1} \rm M_{\odot}$.
This reveals that the clustering strength of low-mass haloes
around high mass haloes (i.e. $M_{\rm vir} \geqslant 10^{11} h^{-1} 
\rm M_{\odot}$) decreases with increasing $M_{\rm cut}$, although the 
dependence on $M_{\rm cut}$ does not appear to be straightforward. In the TruncA
and TruncB runs, we find that $N(r)/N(r)_{\rm CDM}$ is close to unity
out to $r \simeq 10 h^{-1}\,\rm Mpc$, never deviating by more than $10\%$
to within $\sim 500 h^{-1} \rm kpc$ at all redshifts. For the TruncC
and TruncD runs, the suppression in clustering strength is quite marked
-- by $\sim 40\%$ for the TruncC run and $\sim 50\%$ for the TruncD
run. Large deviations at small radii reflect the small numbers of very
close pairs. 
In contrast, the clustering strength of massive haloes (i.e. $M
\geqslant 10^{11} h^{-1} \rm M_{\odot}$) does not appear to be affected
by $M_{\rm cut}$, as we inferred from Figure~\ref{fig:slices}. The
ratio $N(r)/N(r)_{\rm CDM}$ is noisy -- reflecting the lower number
density of massive haloes -- but it is approximately unity between $0
\lesssim z \lesssim 3$.

\section{Mass Accretion \& Merging Histories}
\label{sec:res_accretion}

Suppressing small scale power at early times leads to a reduction in the
clustering of low-mass haloes around massive haloes 
($M_{\rm vir} \gtrsim 10^{11} h^{-1} \rm M_{\odot}$) at $z \lesssim 3$, which 
implies that the number of likely minor mergers a typical halo will experience 
during a given period should decline with increasing $M_{\rm cut}$. We expect 
this to depend on 
both halo mass and epoch. At a given $z$, the merging history of haloes 
with masses $M_{\rm vir} \sim M_{\rm cut}$ should be more
sensitive to the clustering of small scale structure than haloes with
masses $M_{\rm vir} \gg M_{\rm cut}$. Similarly, at
earlier times when the typical collapsing mass $M^{*}$ is smaller and
$M_{\rm cut}$ is a larger fraction of $M^{*}$, we would expect 
the effect of suppressing small scale structure to be more pronounced.

When computing mass accretion and merging rates, we use merger trees for 
all haloes between $5 \times 10^{10} h^{-1} \rm M_{\odot}$ ($\sim 1600$ 
particles) and $10^{13} h^{-1} \rm M_{\odot}$ at $z$=0. Note that we have a 
hard lower limit of $100$ particles for a halo to be retained in our 
catalogues; this corresponds to a mass of $\sim 3.2 \times 10^9 h^{-1} 
\rm M_{\odot}$, and so we cannot identify minor mergers with mass ratios of 
less than $\sim 6\%$ in our most poorly resolved haloes.

\begin{figure}
  \centering
  \includegraphics[width=\columnwidth]{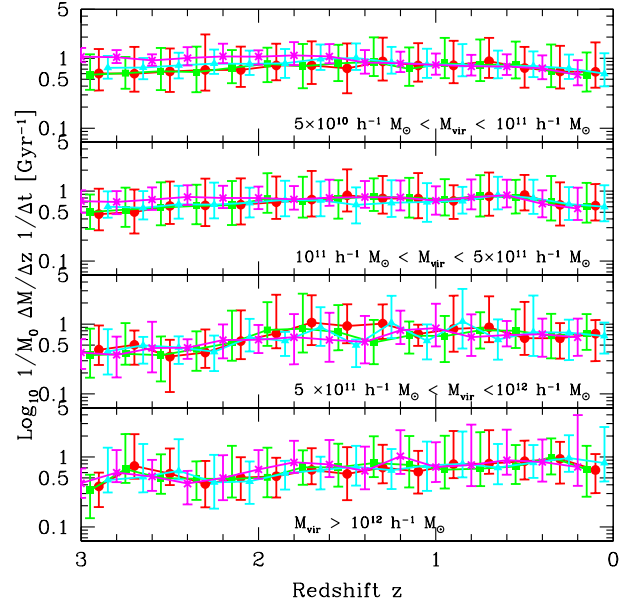}
  \caption{{\bf Impact on Mass Accretion Rate.} For each halo at $z$=0,
    we follow the main branch of its merger tree to higher redshifts
    and compute the difference in virial mass between progenitors at
    $z_0$ and $z_1>z_0$. From this we compute the mass accretion rate
    with respect to time (in Gyrs), normalised by the virial mass of
    the descendent halo at $z$=0. Within each of the mass bins we
    compute the average mass accretion rate for haloes in the fiducial
    $\lambda$CDM run (red filled circles), TruncB ($M_{\rm cut}=10^{10}
    h^{-1} M_{\odot}$; green filled squares),TruncC ($M_{\rm cut}=5
    \times 10^{10} h^{-1} M_{\odot}$; cyan filled triangles) and TruncD 
    ($M_{\rm cut}=10^{11} h^{-1} M_{\odot}$; magenta crosses).}
\label{fig:mass_accretion_rate}
\end{figure}

\subsubsection*{Mass Accretion Rate} 
In Figure~\ref{fig:mass_accretion_rate} we 
show how the mass accretion rate of the most massive progenitors of
haloes identified at $z$=0 evolves with redshift. Note that his accretion rate 
includes both smooth accretion and minor and major mergers. The distinction 
between minor mergers and smooth accretion may be a
moot one in the CDM model -- as the mass and force
resolution of the simulation increases, we continue to resolve 
increasing numbers of low-mass haloes -- but this is not necessarily the case
in the truncated models that we consider. 

From upper to lower panels, we show the average accretion rate as a
function of redshift for haloes with virial masses at $z$=0 in the range 
$5 \times 10^{10} \leqslant M_{\rm vir}/h^{-1} \rm M_{\odot} 
\leqslant 10^{11}$ (filled circles), $10^{11} \leqslant \rm M_{\rm
  vir}/h^{-1} \rm M_{\odot} \leqslant 5 \times 10^{11}$ (filled
squares), $5 \times 10^{11} \leqslant M_{\rm vir}/h^{-1} \rm M_{\odot} 
\leqslant 10^{12}$ (filled triangles) and $M_{\rm vir}/h^{-1} \rm
M_{\odot} \leqslant 10^{12}$ (crosses). Note that we measure the
accretion rate as the change in virial mass ($\Delta M$) per unit
redshift ($\Delta z$) per unit time ($\Delta t$), normalised by the
final (i.e. $z$=0) virial mass. Bars indicate r.m.s. scatter.

Figure~\ref{fig:mass_accretion_rate} shows that haloes accrete their mass 
at similar rates across the different models, regardless of whether or 
not small scale power is suppressed at early times. On average, less massive haloes
tend to have higher accretion rates at $z \gtrsim 1$ than their more 
massive counterparts, but this rate starts to drop $z \sim 1$ and
declines steadily to $z$=0 (see also Figure~\ref{fig:direct_comparison_plots}
for detailed mass accretion histories for individual haloes). In contrast, 
more massive haloes accrete 
their mass at a steady rate. We find 
that our accretion rates for $\Lambda$CDM haloes are in good agreement 
with those consistent with those of, for example, \citet{maulbetsch07}.

\begin{figure}
  \centering
  \includegraphics[width=\columnwidth]{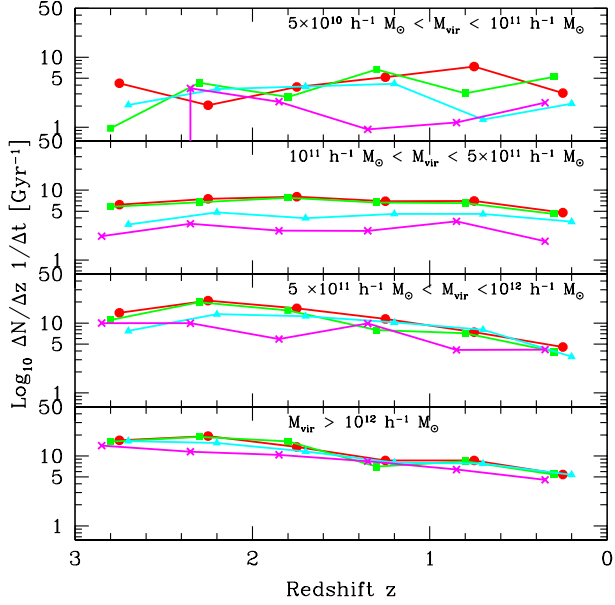}
  \caption{{\bf Impact on Merger Rate.} For each halo at $z$=0,
    we follow the main branch of its merger tree to higher redshifts
    and determine the number of mergers with mass ratios in excess of
    $6\%$ experienced by the halo between $z_0$ and $z_1>z_0$. From 
    this we compute the merger rate per unit redshift per unit time.
    Within each of the mass bins we compute the average merger rate for 
    haloes in the fiducial $\lambda$CDM run (red filled circles), 
    TruncB ($M_{\rm cut}=10^{10} h^{-1} \rm M_{\odot}$; green filled
    squares), TruncC ($M_{\rm cut}=5 \times 10^{10} h^{-1} \rm M_{\odot}$; 
    cyan filled triangles) and TruncD ($M_{\rm cut}=10^{11} h^{-1} 
    \rm M_{\odot}$; magenta crosses).}
\label{fig:merger_rate}
\end{figure}

\subsubsection*{Merger Rates} 
In Figure~\ref{fig:merger_rate}, we focus on the 
merger rate 
${\Delta N/\Delta z/\Delta t}$ and its variation with redshift, where 
$\Delta N$ is the number of mergers per unit redshift per unit time. Here 
differences between runs are immediately apparent and in the sense 
that we expect -- for halo masses close to $M_{\rm cut}$ increases, 
the merging rate decreases. Note that the estimated merger rate is quite noisy in the 
lowest mass bin (upper panel), especially at early times -- in this
case, the lower limit of 100 particles imposed by our halo
catalogues corresponds to a merger of progressively greater mass ratio
with increasing redshift. For this reason, we focus on haloes with
masses at $z$=0 in excess of $10^{11} h^{-1} M_{\odot}$. For haloes
with masses between $10^{11} \leqslant M_{\rm vir}/h^{-1} \rm
M_{\odot} \leqslant 5 \times 10^{11}$, we find that the average merger
rate in the TruncC (TruncD) model is a factor of $\sim 3 (1.5)$ smaller
than that in the fiducial $\Lambda$CDM model, and this is approximately 
constant with redshift. The difference is less pronounced for haloes
with masses between $5 \times 10^{11} \leqslant M_{\rm vir}/h^{-1} 
\rm M_{\odot} \leqslant 10^{12}$, and for haloes with masses in excess
of $10^{12} h^{-1} \rm M_{\odot}$ there is no discernible difference in
the merging rates with redshift.\\

\begin{figure}
  \centering
  \includegraphics[width=\columnwidth]{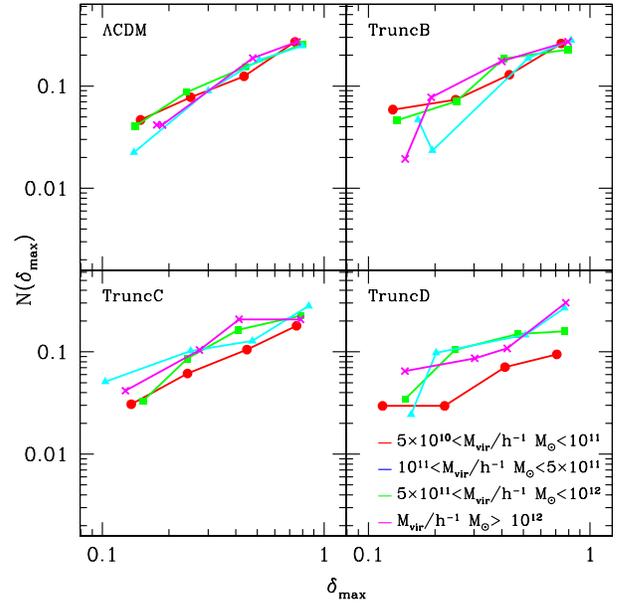}
  \caption{{\bf Distribution of Most Significant Mergers.} For each
  halo at $z$=0, we compute the mass ratio of the most significant
  merger $\delta_{\rm max}$ that it has experienced since $z \simeq$ 0.5 and 
  construct the frequency distribution of $\delta_{\rm max}$ for the
  respective models.}
\label{fig:msigmergers}
\end{figure}

In Figure~\ref{fig:msigmergers} we assess how major mergers are affected by 
suppression of small scale power at early times. This demonstrates that the
likelihood that the mass ratio of the most significant merger
experienced by a halo since $z \simeq 0.5$ does not depend strongly
on whether or not small scale structure has been suppressed. Here we 
follow \citet{power11} and compute the distribution of the most significant 
merger $\delta_{\rm max}=M_{\rm acc}/M_{\rm vir}$ experienced by each halo
(identified at $z$=0) since $\sim 0.5$\footnote{This redshift interval
corresponds to $\sim$2 dynamical times; see \citet{power11}.}, split 
according to virial mass at $z$=0. 

There are a number of interesting trends in this Figure.
The first is that most significant mergers with large mass ratios 
(i.e. minor mergers) are relatively uncommon; the probability
distribution increases approximately as a power law with $\delta_{\rm max}$ as 
$\delta_{\rm max}^{1.2}$. The second is that, in the CDM model, the 
likelihood that a halo experiences a most significant merger with a
given $\delta_{\rm max}$ does not depend strongly on its mass. For
example, a halo with virial mass of $10^{11} h^{-1} \rm M_{\odot}$ is as 
likely to have experienced a major merger with mass ratio of $\sim
50\%$ as a $10^{13} h^{-1} M_{\odot}$ halo -- approximately $20\%$. The
third is that there is some evidence that haloes in the mass range 
$5 \times 10^{10} \leqslant M_{\rm vir}/h^{-1} \rm M_{\odot} 
\leqslant 10^{11}$ are less likely to experience major mergers with
mass ratios in excess of $\sim 50\%$ (compare TruncB and TruncD).

\section{Angular Momentum Content}
\label{sec:res_spin_growth}

Suppressing small scale power at early times impacts on both the clustering strength 
of low-mass haloes and the rate at which mergers and accretions at later times. Do 
we see a corresponding influence on the angular momentum content of haloes at later
times?

\subsubsection*{Spin Parameter} 
We begin by considering the spin parameter $\lambda$,
which quantifies the degree to which the halo is supported by rotation and which
we define using the ``classical'' definition of \citet{peebles69},
\begin{equation}
  \label{eq:lambda}
	{\lambda = \frac{J|E|^{1/2}}{GM_{\rm vir}^{5/2}}.}
\end{equation}

\noindent Here $J$ and $E$ are the total angular momentum and binding energy
respecively of material with $r_{\rm vir}$ and $G$ is the gravitational 
constant. We impose a lower limit of 600 particles within $r_{\rm vir}$ 
($M_{\rm vir} \geqslant 2\times10^{10} h^{-1}\,\rm M_{\odot}$) when 
measuring $\lambda$; this ensures that both $J$ and $E$ are unaffected 
by discreteness effects \citep[cf.][]{power11}.

\begin{figure}
  \centerline{\includegraphics[width=\columnwidth]{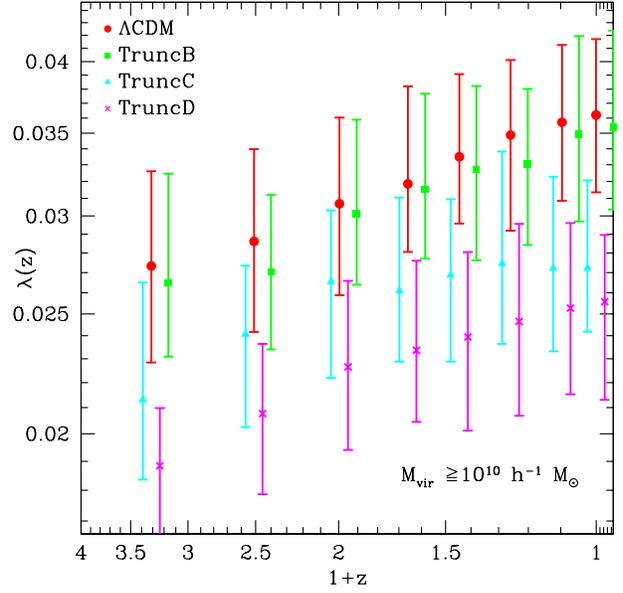}}
  \centerline{\includegraphics[width=\columnwidth]{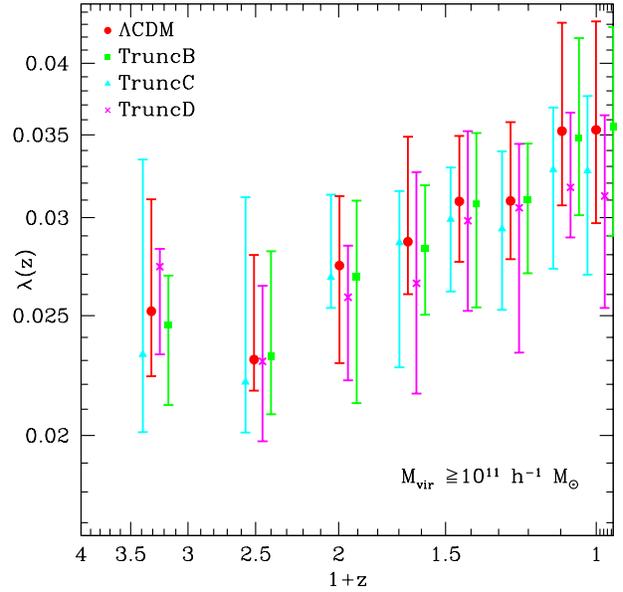}}
  \caption{{\bf Variation of Median $\lambda$ with Redshift.} We
    show how the median spin parameter $\lambda_{\rm med}$
    varies with redshift. In the left hand panel we consider
    all haloes with virial masses in excess of $M_{\rm vir}
    \geqslant 1.9 \times 10^{10} h^{-1} M_{\odot}$, while in the
    right hand panel we consider all haloes that satisfy 
    $M_{\rm vir} \geqslant 10^{11} h^{-1} M_{\odot}$. Lower and upper
    error bars represent the $25^{\rm th}$ and $75^{\rm th}$
    percentiles. The filled circles, squares, triangles and crosses
    correspond to the fiducial $\Lambda$CDM, TruncB, TruncC and
    TruncD runs respectively.}
  \label{fig:spin_median_growth}
\end{figure}

In Figure~\ref{fig:spin_median_growth} we show how the median spin of 
the halo population evolves with redshift. In the upper panel we
focus on the haloes with $M_{\rm vir} \geqslant 2 \times 10^{10} h^{-1} M_{\odot}$,
while in the lower panel we consider haloes with $M_{\rm vir} \geqslant 10^{11}
h^{-1} M_{\odot}$. Filled circles, squares, triangles and crosses
represent the median spin of the halo populations in the $\Lambda$CDM, TruncB,
TruncC and TruncD runs, and error bars indicate the 25$^{\rm th}$ and 
75$^{\rm th}$ percentiles of the distribution. This figure suggests that
the behaviour of the distribution of $\lambda$ is sensitive to $M_{\rm cut}$
-- systematic differences are apparent in the TruncC and TruncD runs
when we include all haloes with $M_{\rm vir} \geqslant 2 \times 10^{10} h^{-1}
M_{\odot}$, whereas the distributions are statistically similar when we
include only haloes with $M_{\rm vir} \geqslant 10^{11} h^{-1} M_{\odot}$. 

This figure is interesting because we include a large population of haloes in 
the TruncC and TruncD runs with $M_{\rm vir} \leq M_{\rm cut}$ when we include haloes with 
$M_{\rm vir} \geqslant 2 \times 10^{10} h^{-1} M_{\odot}$, and so the apparent
differences are to be expected. In contrast, we do not see any evident 
differences when we include haloes with $M_{\rm vir} \geqslant 10^{11} h^{-1}
M_{\odot}$. This is also interesting because it reveals 
that the median $\lambda$ increases with decreasing redshift at approximately 
the same rate -- in proportion to $(1+z)^{-0.3}$ -- regardless of whether or 
not we include haloes with masses below $M_{\rm cut}$.\\

\begin{figure}
  \includegraphics[width=\columnwidth]{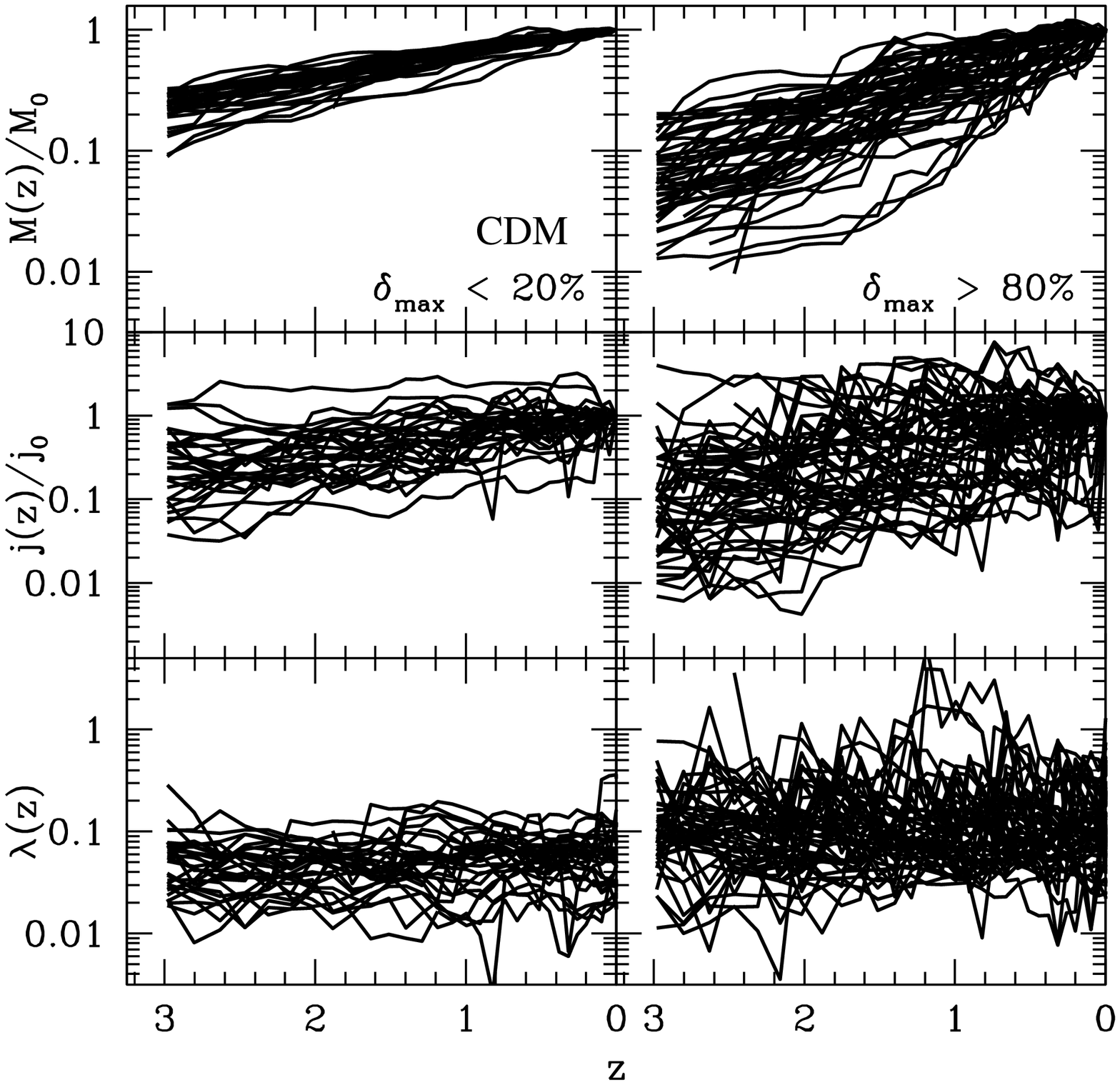}
  \includegraphics[width=\columnwidth]{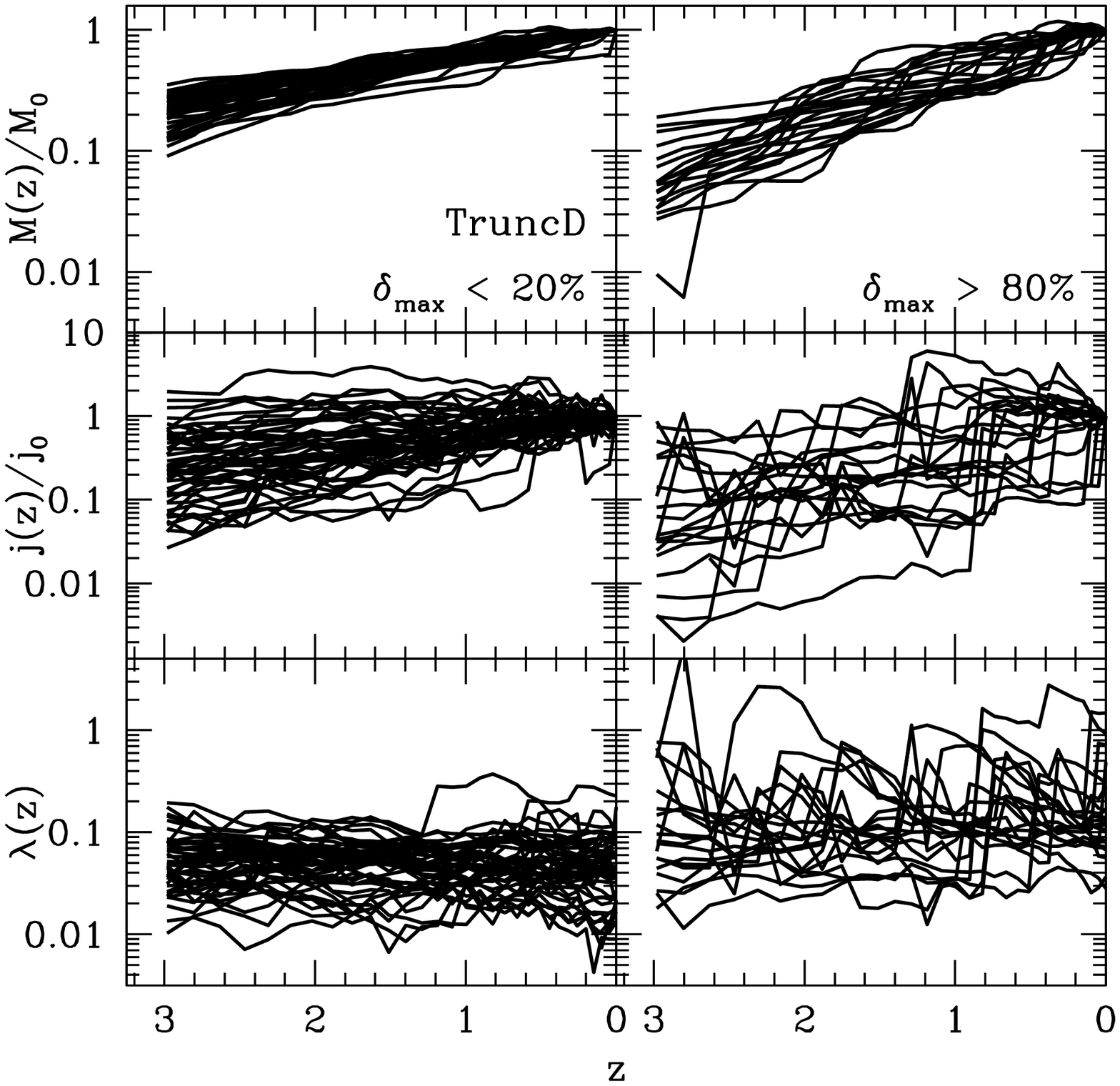}
  \caption{{\bf Variation of $\lambda$ and $j$ with Redshift
      for Relaxed and Unrelaxed Haloes.} We use the merging histories
      of haloes to identify two samples of haloes, one with a quiescent
      merging history ($\delta_{\rm max} \lesssim 0.2$ since $z$=3.0;
      left hand panels) and one with a violent merging history 
      ($\delta_{\rm max} \gtrsim 0.8$ over the same period; right hand
      panels) in the $\Lambda$CDM and TruncD runs (upper and lower panels respectively).
      Haloes are chosen such that their virial mass at $z$=0 
      satisfies $M_{\rm vir} \geqslant 10^{11} h^{-1} M_{\odot}$ ($\sim 3000$
      particles). The upper, middle and lower panels show the growth of halo
      virial mass (normalised to the virial mass at $z$=0),
      the specific angular momentum $j=J/M$ (normalised to its value at $z$=0,
      $j_0$) and
      dimensionless spin parameter $\lambda=J|E|^{1/2}/GM_{\rm vir}^{5/2}$ as a
      function of redshift $z$. }
\label{fig:spin_jm_mah}
\end{figure}

In Figure~\ref{fig:spin_jm_mah} we focus on individual haloes, showing 
how $\lambda$ and the specific angular momentum $j=J/M$ vary with redshift 
$z$ for a selection of haloes with
quiescent and violent merging histories, drawn from haloes with $M_{\rm vir}
\geqslant 10^{11} h^{-1} \rm M_{\odot}$ over the redshift interval 
$0 \leq z \leq 3$. For each halo we determine the most significant merger 
$\delta_{\rm max}$ that it has experienced since $z$=1, where we define 
$\delta_{\rm max}$ as the mass ratio of the most major merger experienced 
by the main progenitor of a halo identified at $z$=0 during the redshift 
interval $0 \leq z \leq 1$ \citep[cf.][]{power11}.
This gives a distribution of $\delta_{\rm max}$ and we identify haloes in the 
upper (lower) $20\%$ of the distribution as systems with violent (quiescent) 
merging histories. For ease of comparison, we focus on the extremes -- the 
$\Lambda$CDM and TruncD runs (top and bottom respectively).

There are a few of points worthy of note in relation to the evolution of the
spin parameter with redshift. First, the spin parameter for
a given halo is a very noisy quantity but if we consider the average
behaviour of haloes in the respective samples, we do not find any clear
correlation between spin and redshift (based on their Spearman rank
coefficient). Second, there is a clear offset 
between median spins in the quiescent and violent samples --
haloes with violent merging histories tend to have higher spins (by 
factors of $\sim 3$-$4$) than haloes with quiescent histories. However, 
there is appreciable scatter over any given halo's history -- the
r.m.s. variation is $\sim 0.25$-$0.29$ for haloes in the quiescent
sample and $\sim 0.39$-$0.42$ in the violent sample. Importantly, third, it is
the dynamical state and merging history of a halo that has greater impact
on its instantaneous spin and specific angular momentum -- the influence
of the dark matter is a secondary effect at best.

\begin{figure*}
  \centering
  \includegraphics[width=0.3\textwidth]{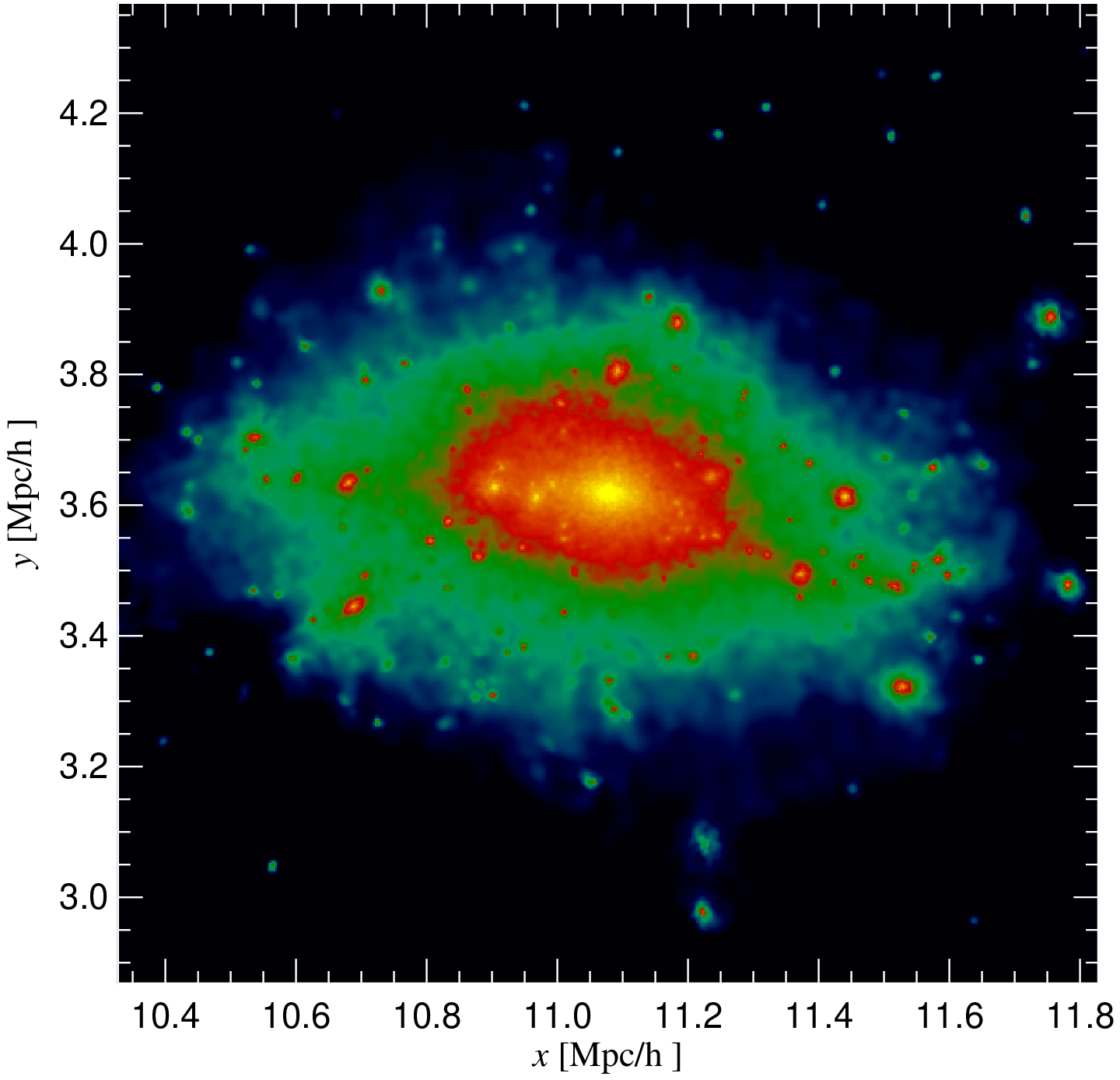}
  \includegraphics[width=0.3\textwidth]{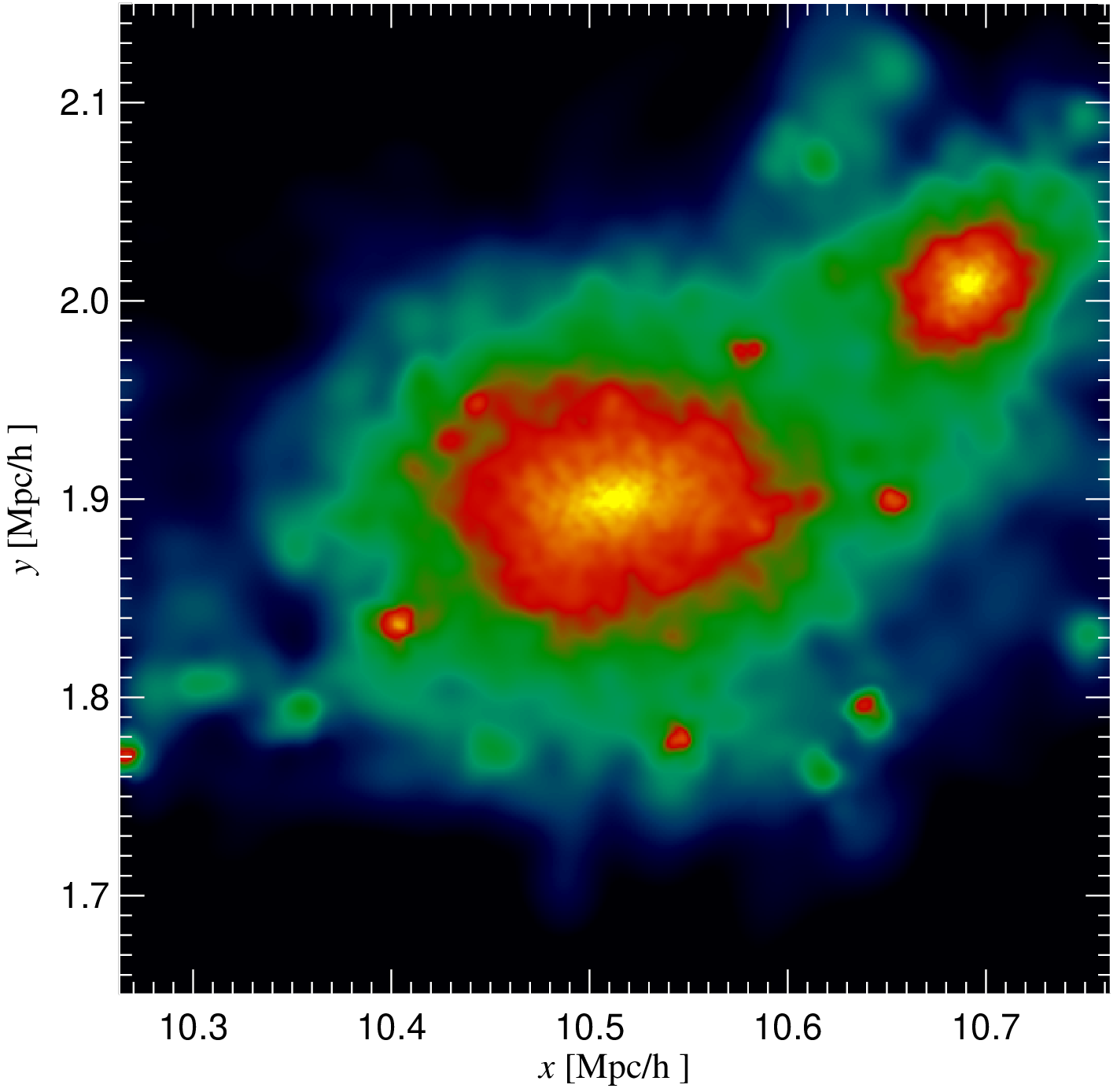}
  \includegraphics[width=0.3\textwidth]{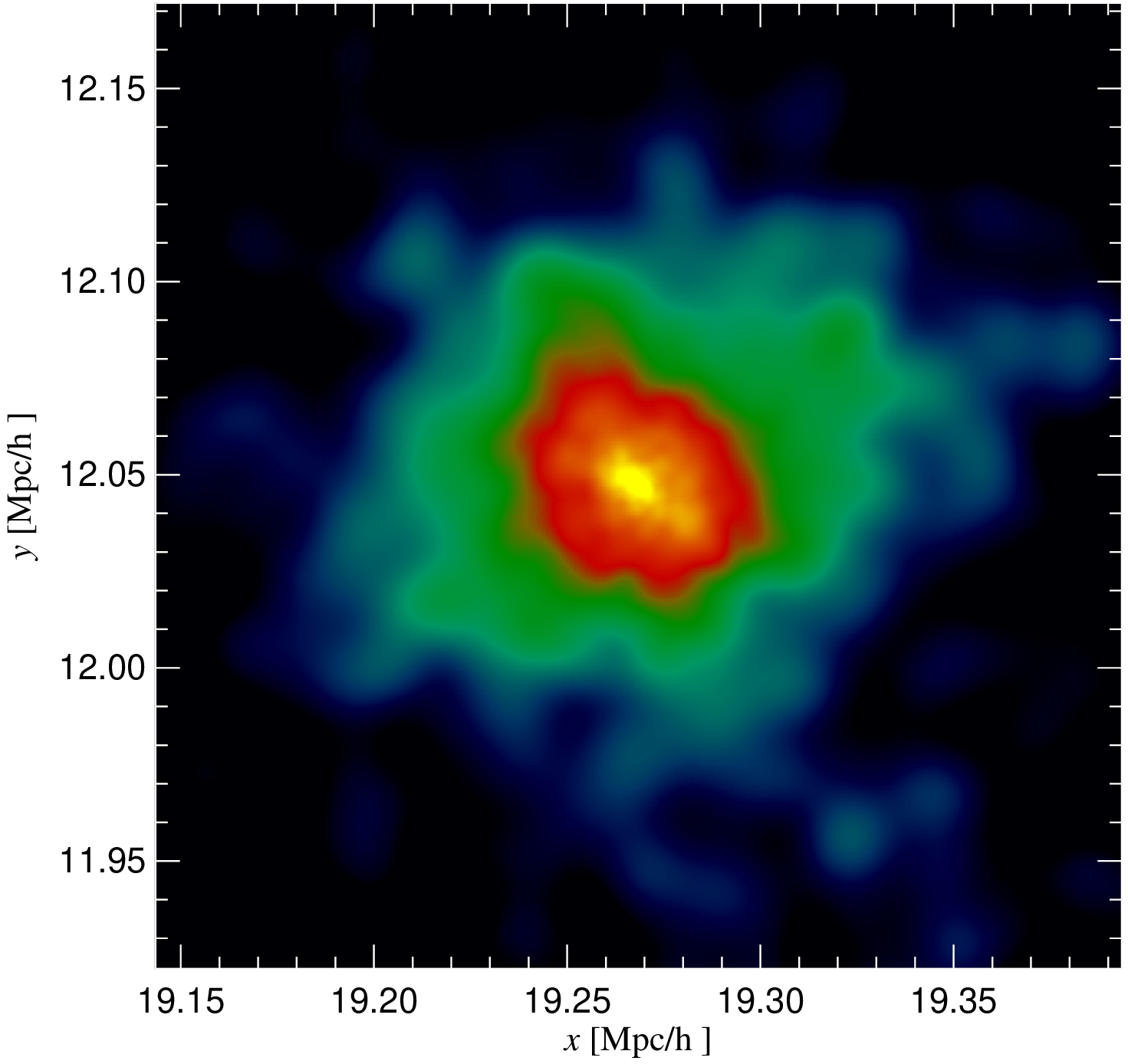}

  \centering
  \includegraphics[width=0.3\textwidth]{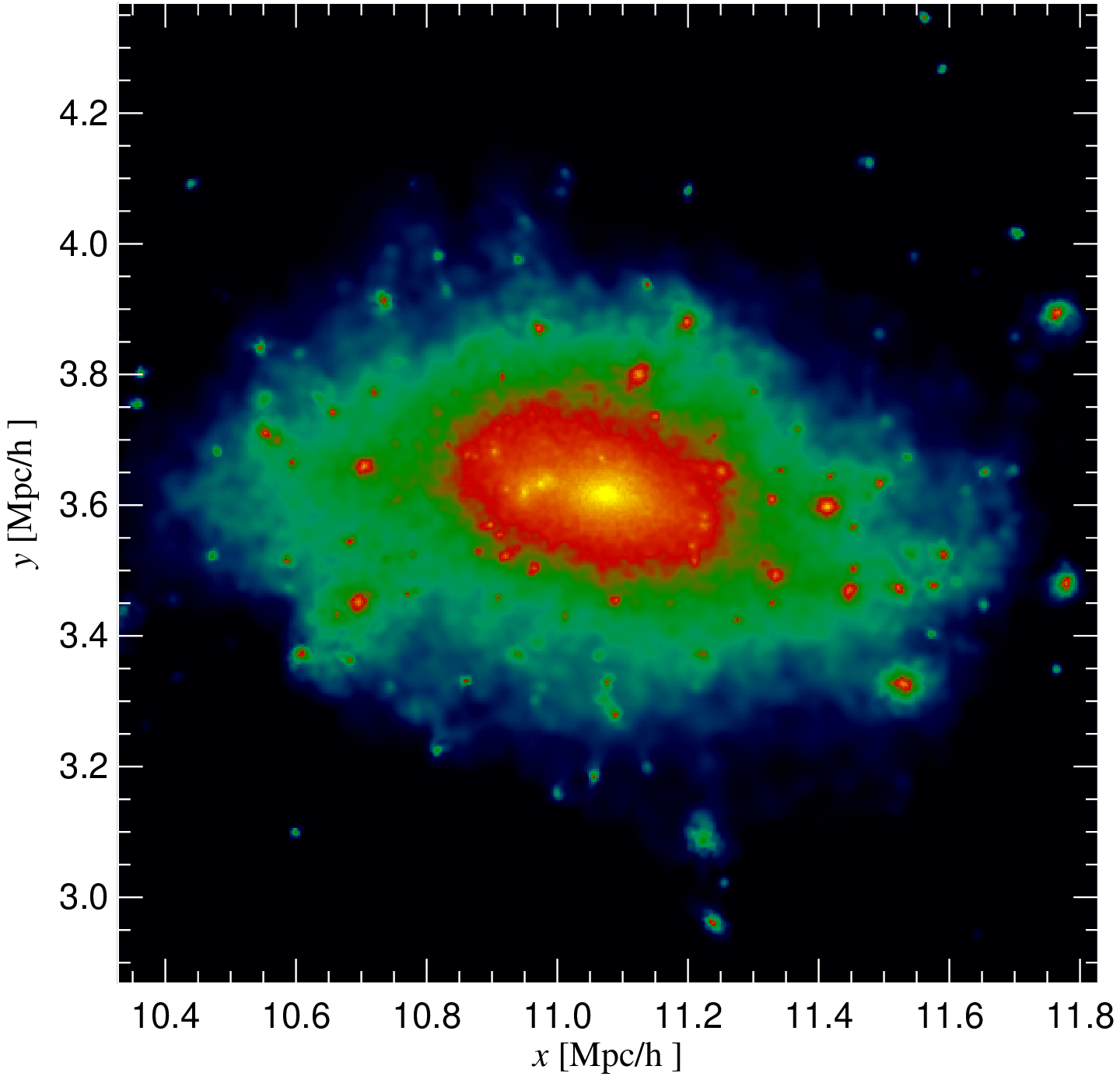}
  \includegraphics[width=0.3\textwidth]{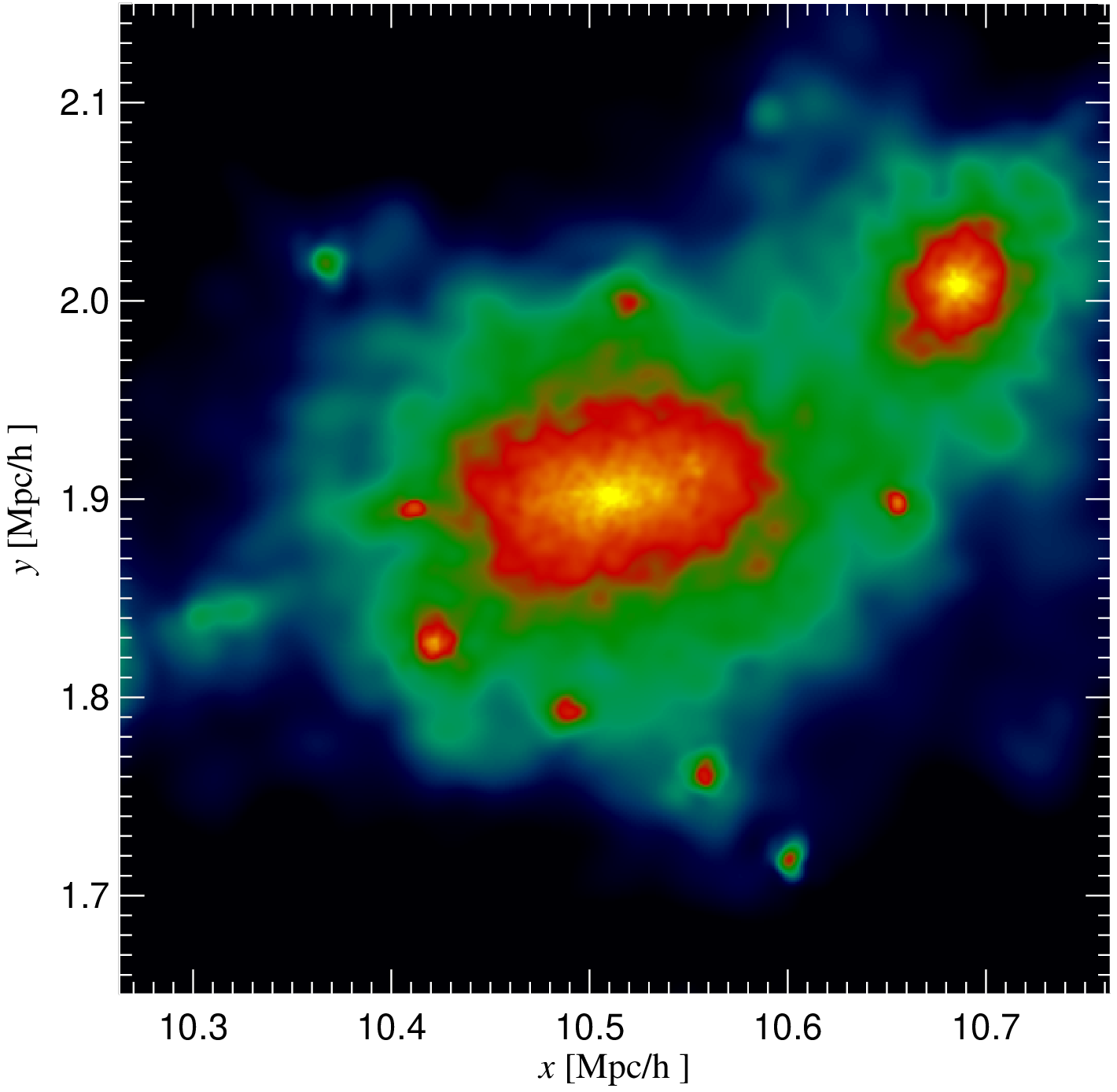}
  \includegraphics[width=0.3\textwidth]{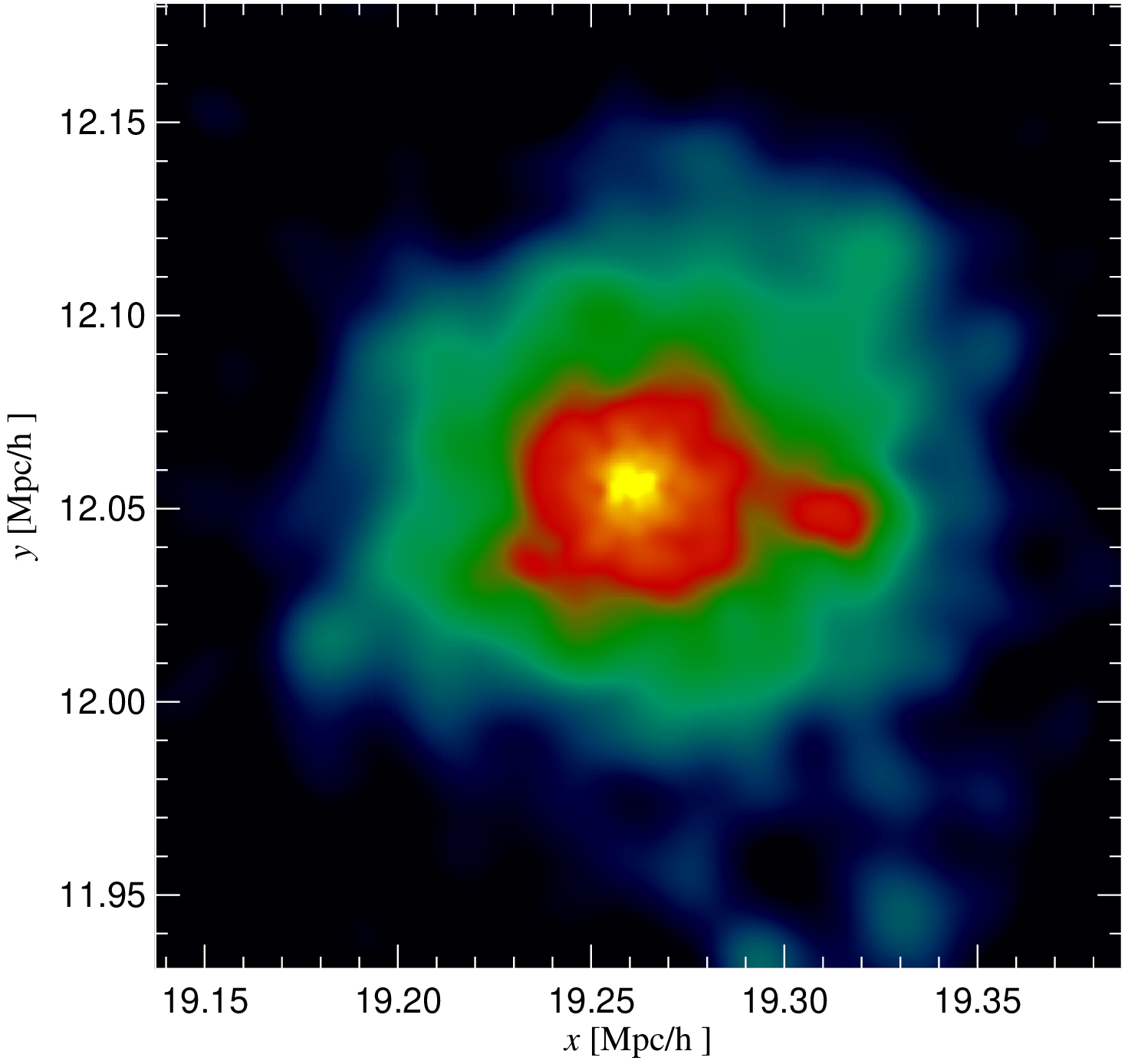}

  \centering
  \includegraphics[width=0.3\textwidth]{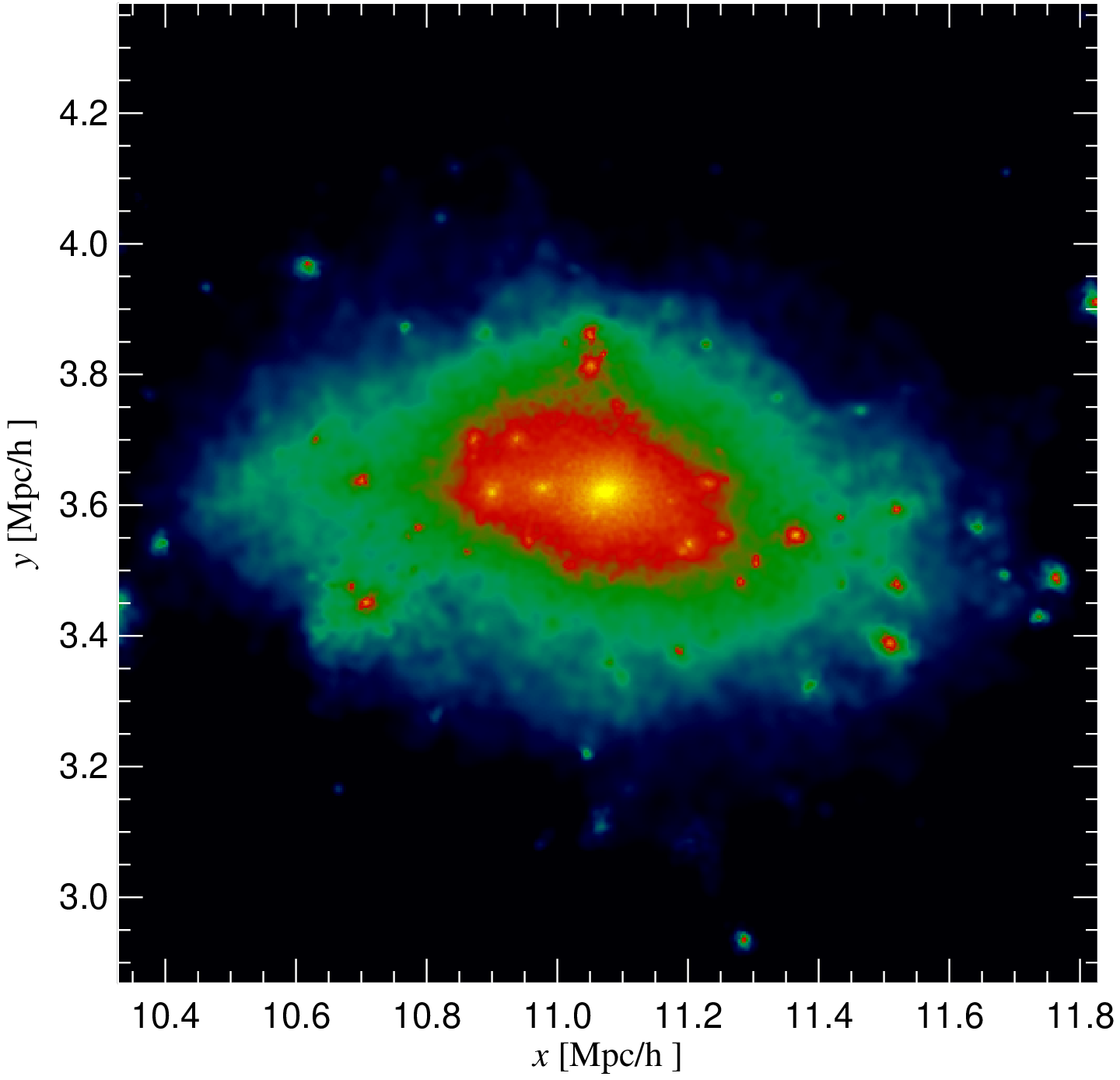}
  \includegraphics[width=0.3\textwidth]{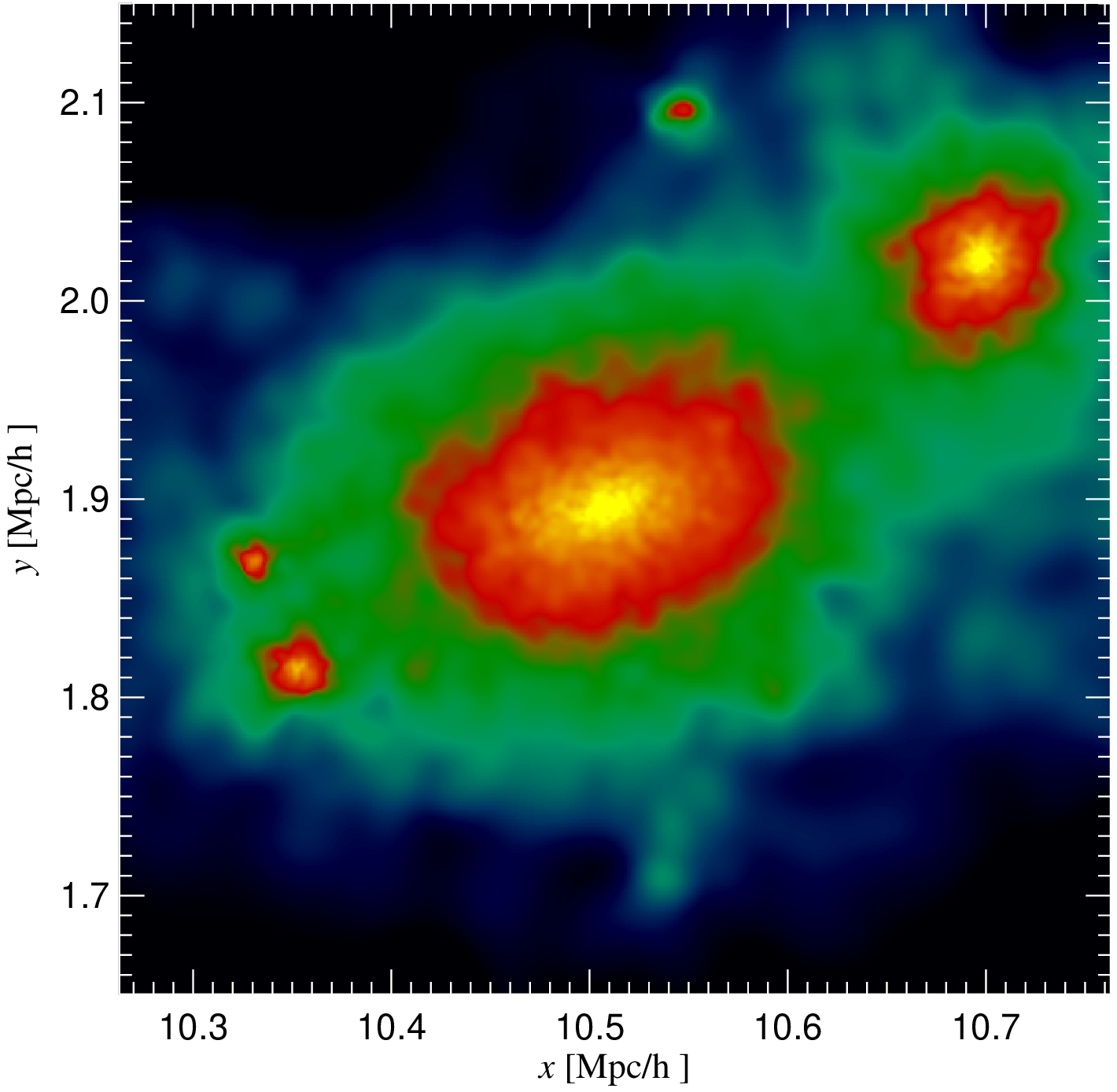}
  \includegraphics[width=0.3\textwidth]{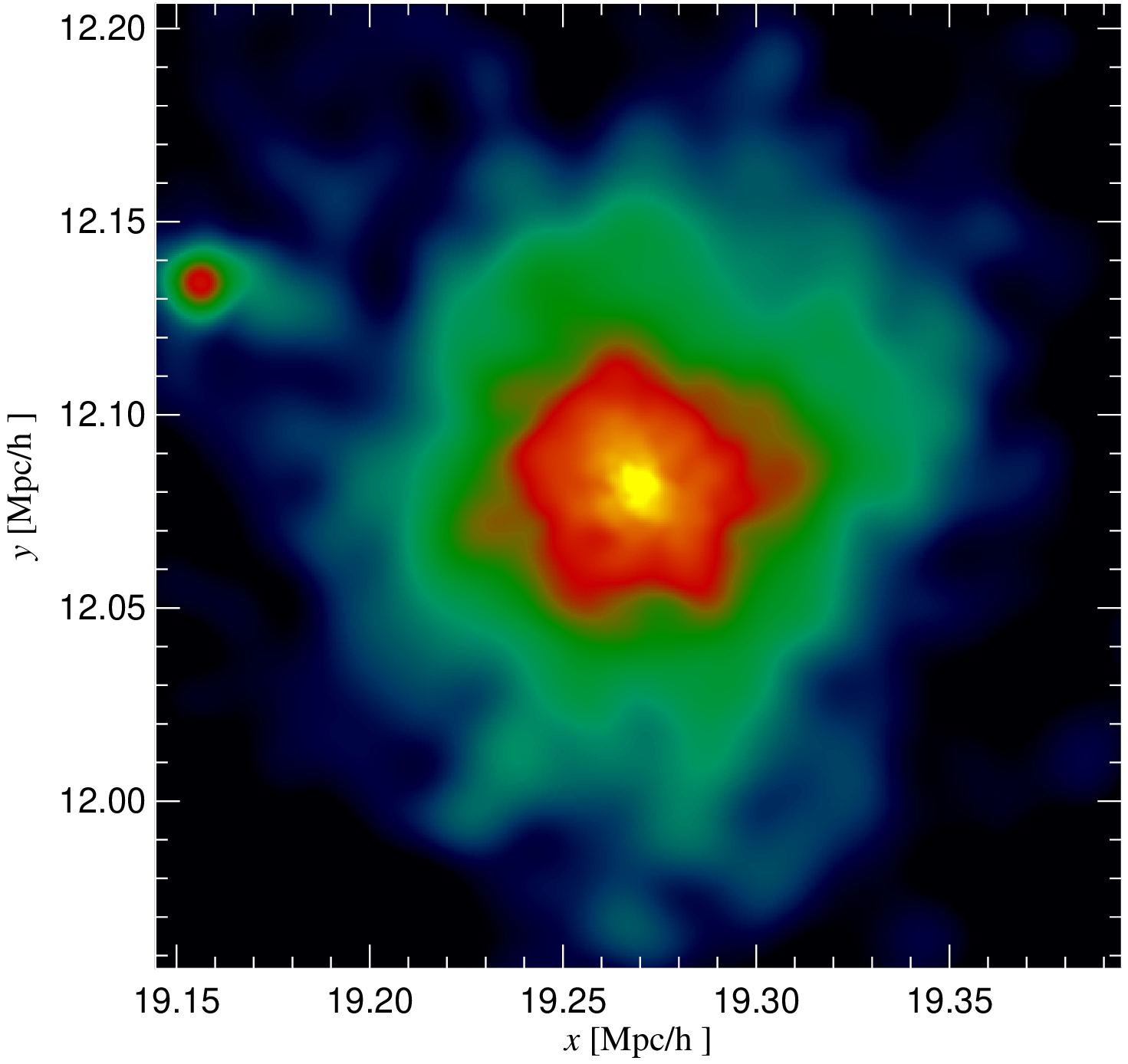}

  \centering
  \includegraphics[width=0.3\textwidth]{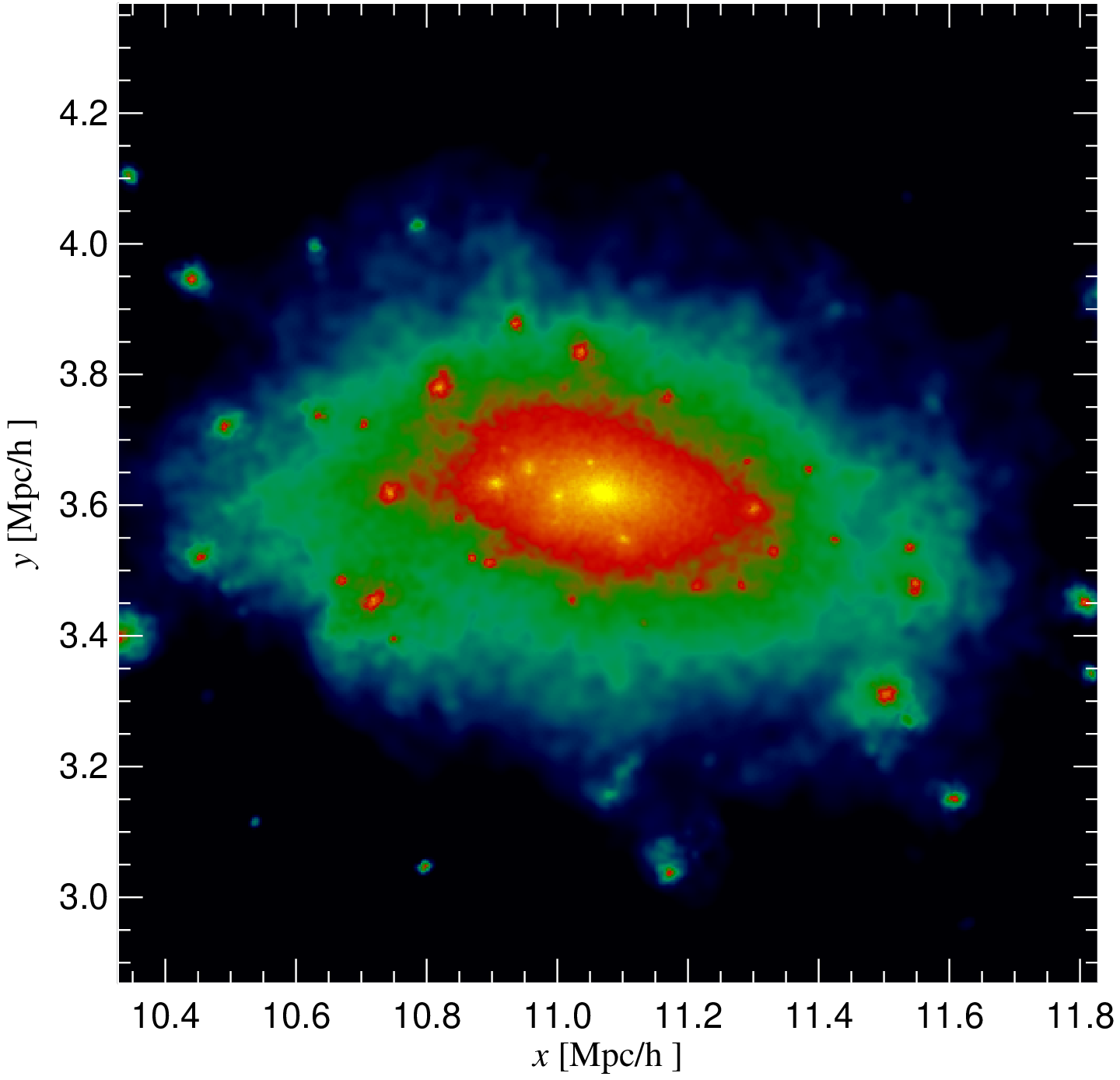}
  \includegraphics[width=0.3\textwidth]{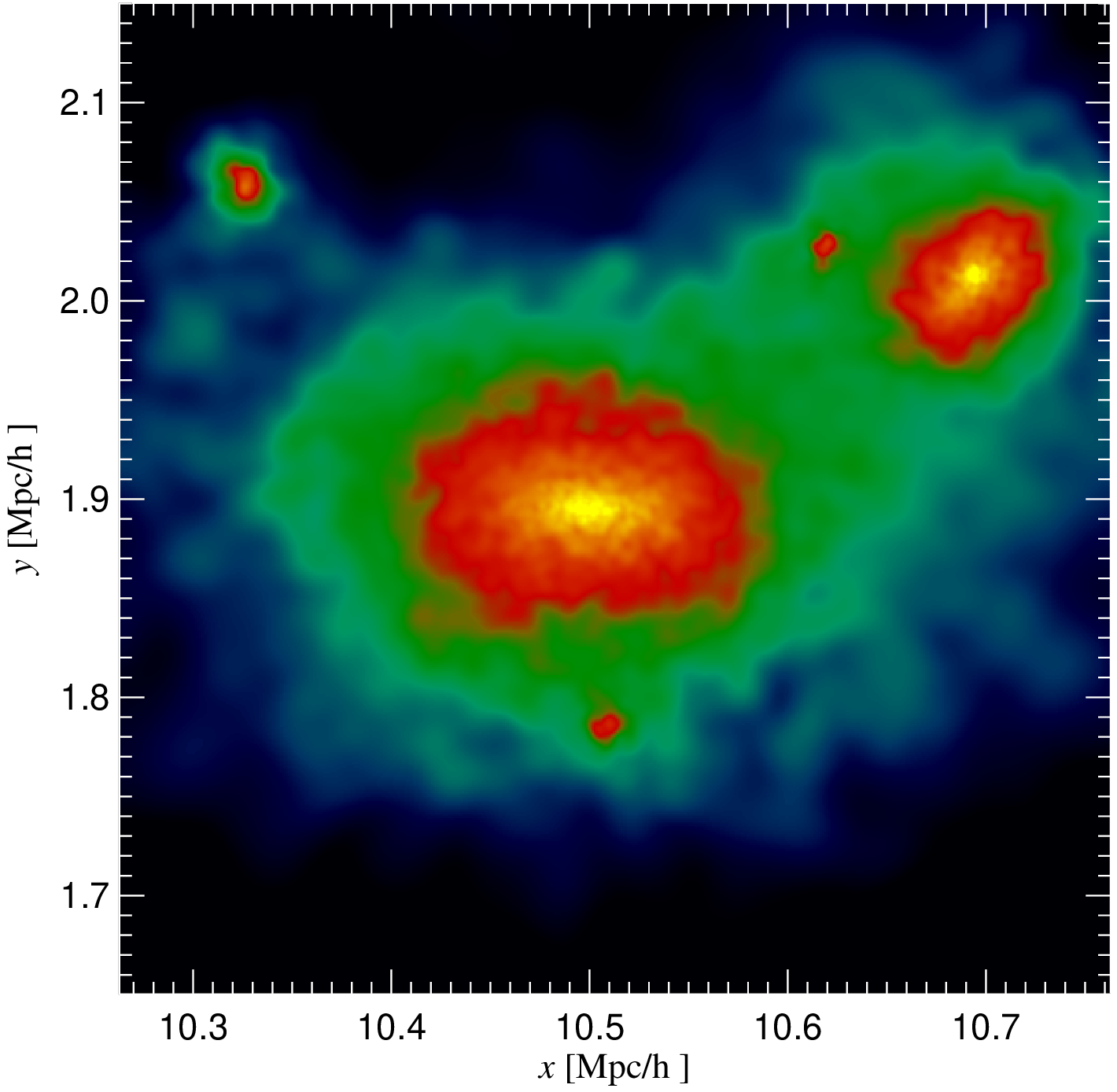}
  \includegraphics[width=0.3\textwidth]{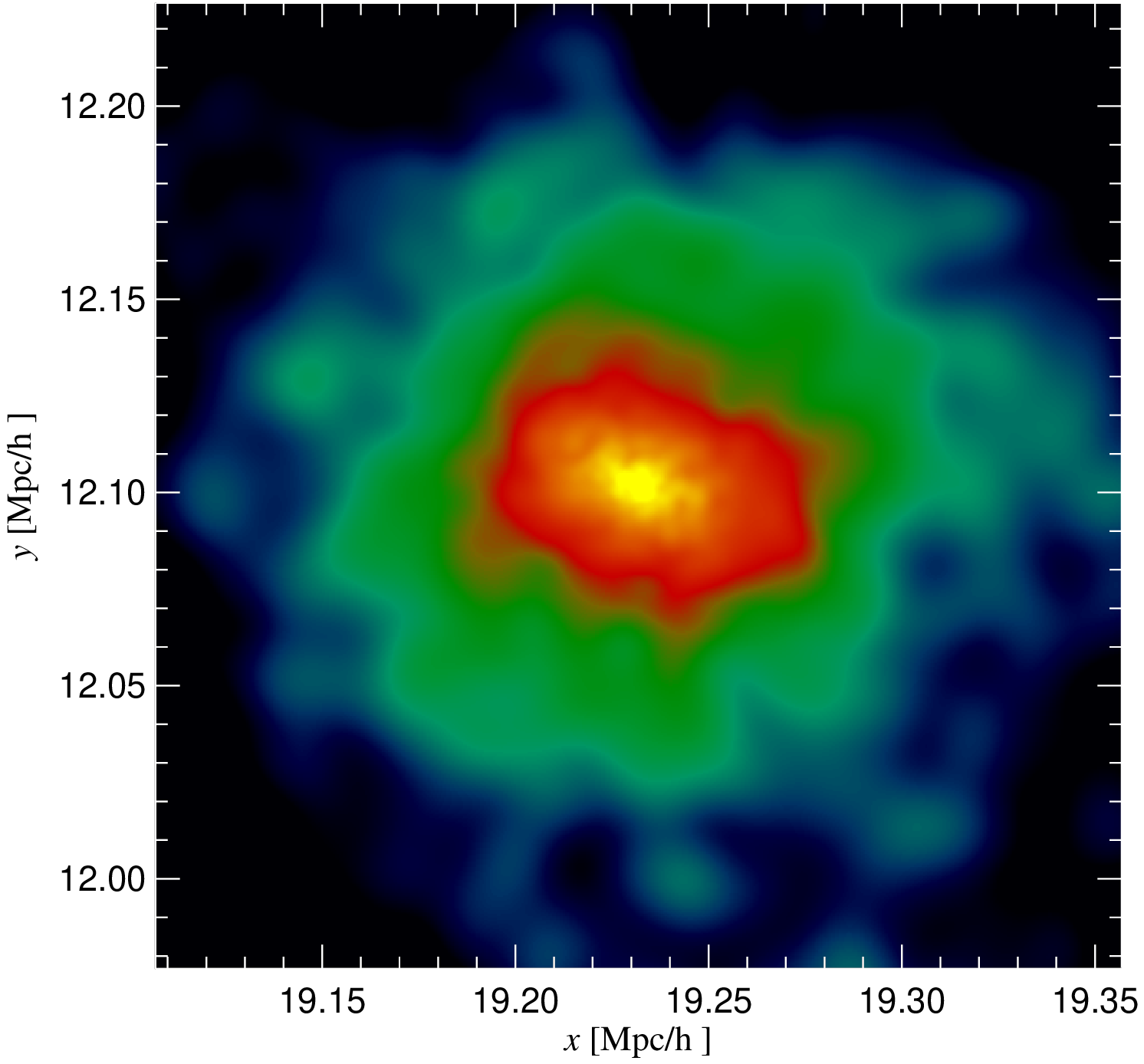}

  \caption{{\bf Direct Comparison of Haloes: Projected Dark Matter Density Maps.}
  From left to right, haloes with virial masses at $z$=0 of $M_{\rm vir} \simeq (7.85,0.61,0.076)
  \times 10^{12} h^{-1} \rm M_{\odot}$ in the CDM, TruncB, TruncC and TruncD (from top to bottom).}
  \label{fig:direct_comparison_images}
\end{figure*}

\begin{figure}
  \centering
  \includegraphics[width=7cm]{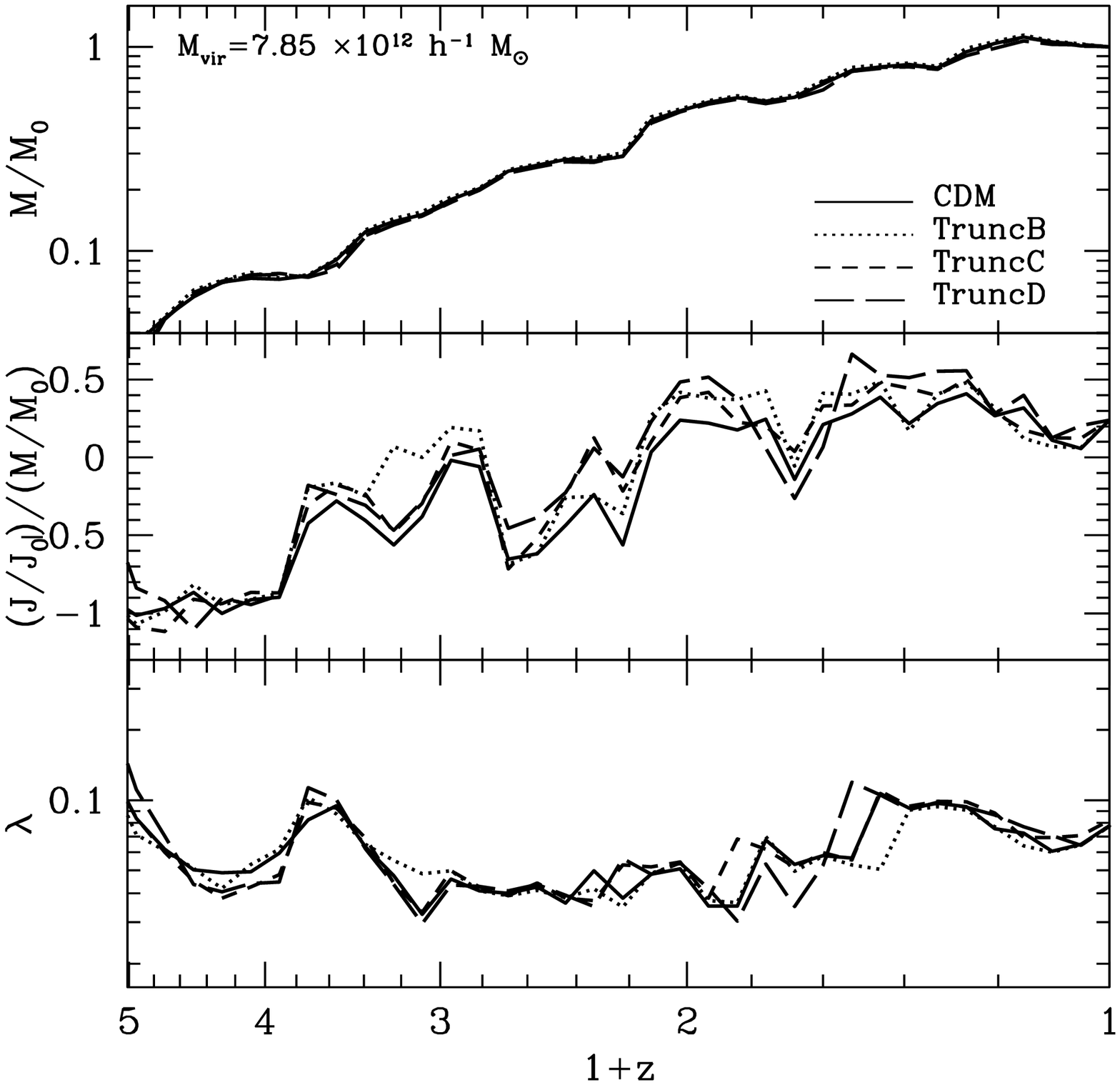}
  \centering
  \includegraphics[width=7cm]{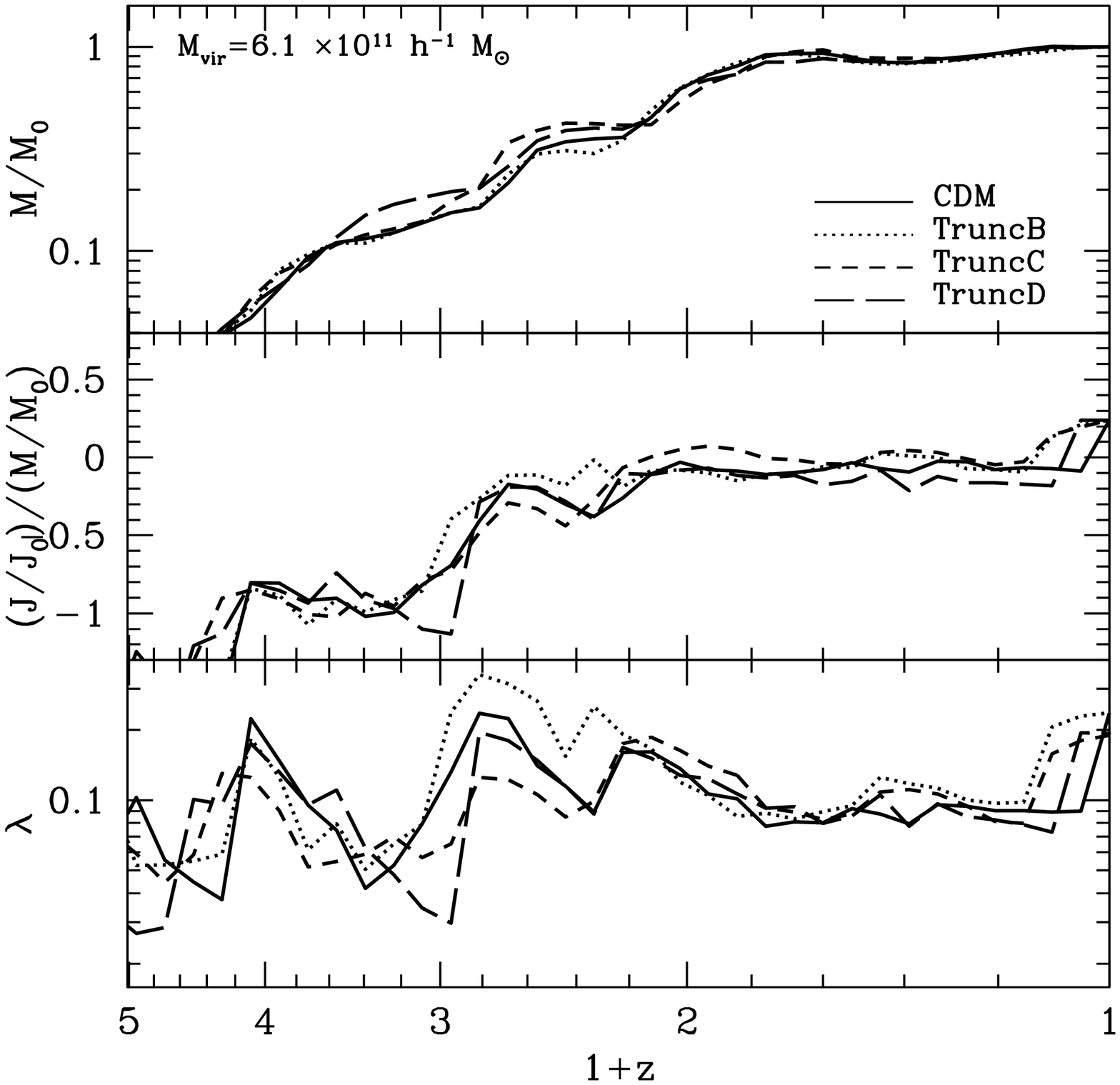}
  \centering
  \includegraphics[width=7cm]{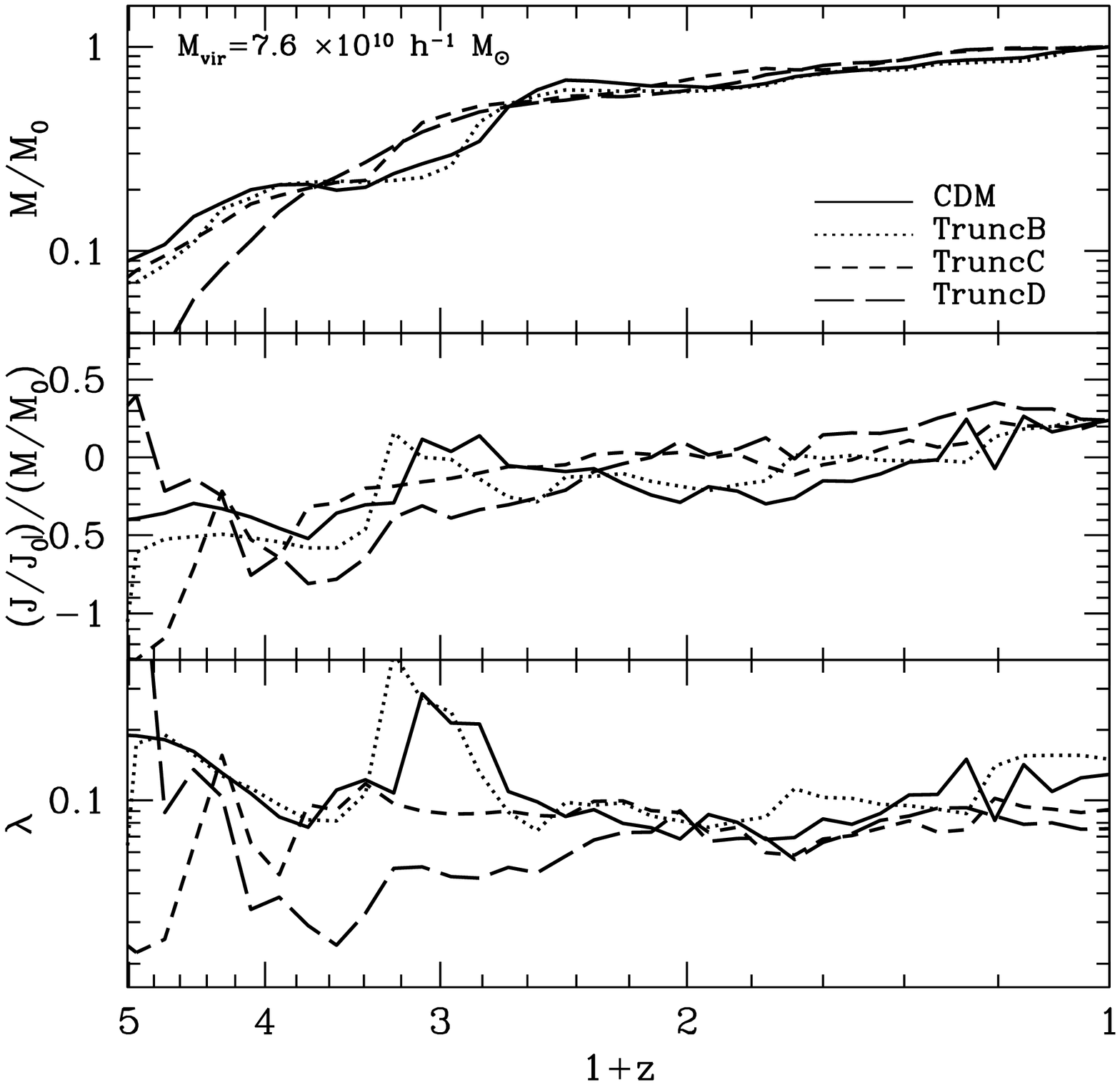}
  \caption{{\bf Direct Comparison of Haloes: Redshift Evolution of Spin and Specific 
      Angular Momentum Evolution.} Upper/middle/lower panel show growth of virial
    mass (normalised to $M_{\rm vir}$ at $z$=0), specific angular momentum (normalised
    to value at $z$=0) and spin parameter $\lambda$ as function of $1+z$.}
  \label{fig:direct_comparison_plots}
\end{figure}

Note that we also compare the growth of angular momentum and spin for three 
sets of cross matched haloes across dark matter models -- shown in 
Figures~\ref{fig:direct_comparison_images} and
\ref{fig:direct_comparison_plots}. From our cross matched catalogues we 
identified blindly a set of three haloes with 
$M_{\rm vir} \simeq (7.85,0.61,0.076) \times 10^{12} h^{-1} \rm M_{\odot}$, 
which are approximately 1, 10 and 100 times the threshold mass of 
$M_{\rm cut}$=$10^{11}\,h^{-1} \rm M_{\odot}$.

Projections of the density distribution in cubes approximately $2\,r_{\rm vir}$ 
on a side and centred on the haloes are shown in 
Figure~\ref{fig:direct_comparison_images} -- 
high, intermediate and low mass haloes (left, middle and right panels) 
in the $\Lambda$CDM, TruncB, TruncC and TruncD models (top to bottom panels 
respectively). Qualitatively the haloes appear similar, with the decreasing
abundance of substructure with increasing severity of truncation in initial 
$P(k)$ the key difference between the models. There are small differences in
the orientation of the intermediate and low-mass haloes (compare, for example,
the intermediate mass halo in the TruncC and TruncD runs) and in the positions
of subhaloes (compare, for example, the low mass halo in the TruncB and TruncD
runs), but such differences are to be expected at the mass and force resolution 
of our simulations.

Figure~\ref{fig:direct_comparison_plots} shows in detail how the virial mass
(upper panels), specific angular momentum (middle panels), and spin parameter
(lower panels) grows over time for each of the three sets of haloes. For the
most massive halo, the mass assembly histories are indistinguishable, while
the specific angular momentum and spin growth are in very good agreement with
each other. For the intermediate mass halo, there are differences in the
mass assembly histories at $z \gtrsim 1$, with the TruncC and Trunc deviating 
from the $\Lambda$CDM and TruncB cases, but they are negligible; the specific angular
momentum and spin growth show small differences but they are in good broad 
agreement. For the low mass halo, it's noticeable that the mass growth is 
in good general agreement across the models at $z \lesssim 3$, but the mass of
the halo in the TruncD case has to grow rapidly to catch up with its 
counterparts in the $\Lambda$CDM, TruncB and TruncC runs at $z \gtrsim 3$. This
has a knock-on effect in the growth of its specific angular momentum and spin
parameter; however, the mass, specific angular momentum and spin parameter
growth are in very good agreement for $z \lesssim 1$.

\begin{figure}
  \includegraphics[width=\columnwidth]{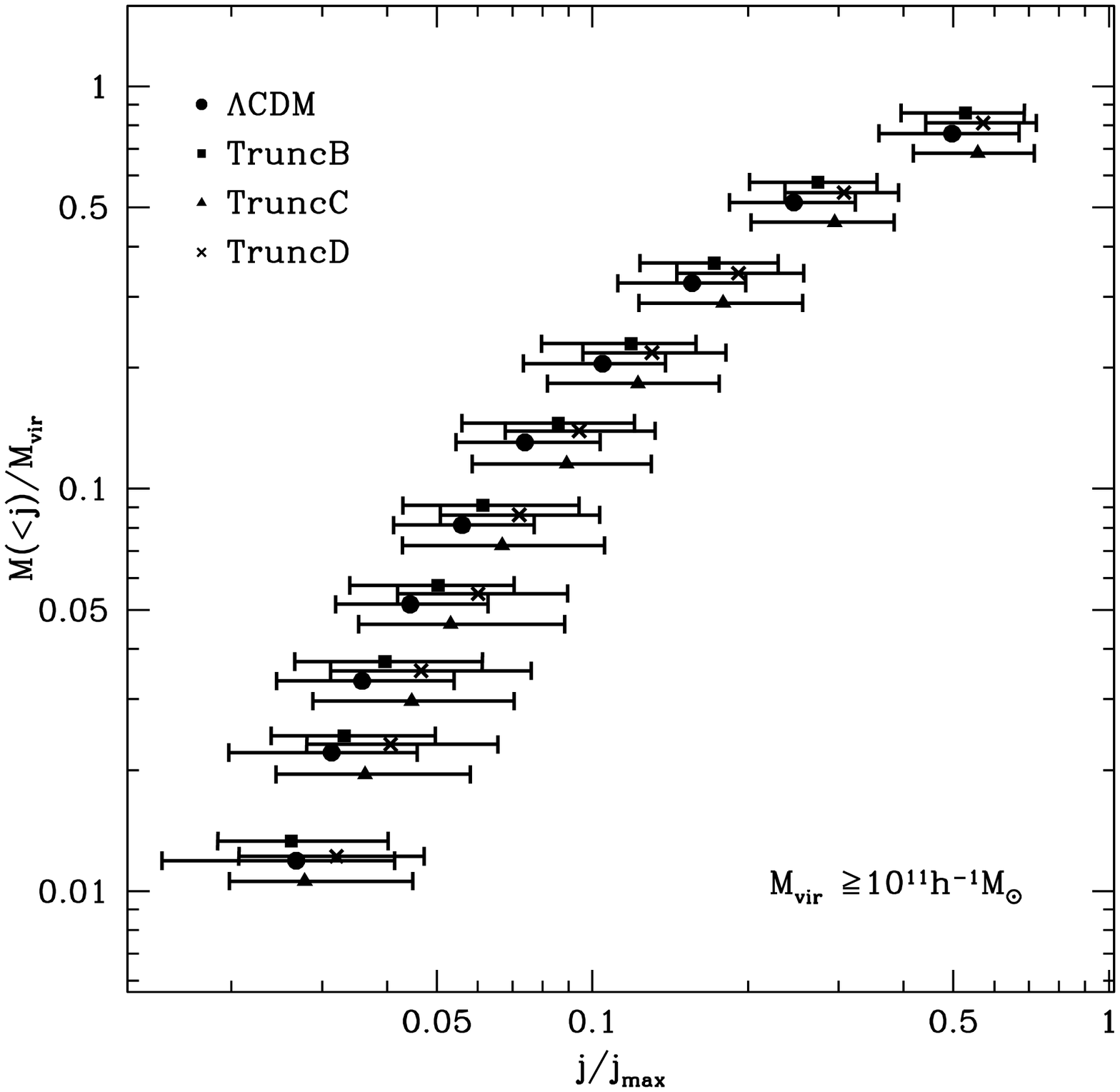}
  \caption{{\bf Specific Angular Momentum Profiles.} We use the method
    of \citet{bullock01a,bullock02} to determine the fraction of halo mass
    that has a total specific angular momentum of $j$ or less. Note that 
    we consider only haloes that satisfy $M_{\rm vir} \geqslant 10^{11} h^{-1} M_{\odot}$.}
  \label{fig:angmom_profiles}
\end{figure}

\subsubsection*{Specific Angular Momentum Profiles} 
There does not appear
to be any systematic difference in the bulk angular momenta of haloes, i.e.
the total angular momentum of material within $r_{\rm vir}$. What of the 
distribution of angular momentum within $r_{\rm vir}$? We focus on the specific 
angular momentum profile, which quantifies the fraction of material within the 
virial radius that has specific angular momentum of $j$ or less. 
Figure~\ref{fig:angmom_profiles} shows the average specific angular
momentum profile $M(<j)$ of haloes in each of our models.

We compute specific angular momentum profiles using the method
presented in \citet{bullock01a,bullock02}. In brief, 
we compute the total angular momentum of the halo and define this 
as the $z$-axis; then we sort particles into spherical shells of 
equal mass and increasing radius, and we assign shell particles to 
one of three equal volume segments determined by the particle's angle with 
respect to the $z$-axis; finally, we compute both the total and 
$z$-component of the specific angular momentum in each segment. This 
allows us to compute the fraction of halo mass with specific angular
momentum of $j$ (and its $z$ component $j_{\rm z}$) or less. Note that
we scale our profiles by $j_{\rm max}$, the maximum specific angular
momentum that we measure in our data; this is distinct from the $j_{\rm
  max}$ used in \citet{bullock01a,bullock02}, who estimate 
$j_{\rm max}$ by fitting their universal angular momentum profile.

In Figure~\ref{fig:angmom_profiles} we show the specific angular momentum 
profile for the total angular momentum $j$, although the $j_z$ behaviour is 
similar. For ease of comparison, we have applied small offsets 
to the data points from the truncated models. There are a few points
worthy of note in this Figure. The profile gently curves towards shallower 
logarithmic slopes with increasing $j$; we find that $M(<j) \sim j^{5/2}$ 
for the lowest angular momentum material and $M(<j) \sim j^{1/2}$ for the 
highest angular momentum material. It is interesting that there is a
systematic trend for lower angular momentum material in the
$\Lambda$CDM and TruncB runs to have \emph{on average} lower values of
$j$ than the TruncC and TruncD runs -- the difference is of order
$25\%$ at most. This trend is not evident when one looks at the 
projected specific angular momentum ($j_z$) profile. However, the
r.m.s. scatter is large for a given $M(<j)$ or $M(<j_{\rm z})$ in all
our models, and for interesting values of $M_{\rm cut} \sim 10^9 h^{-1}
M_{\odot}$ (comparable to our TruncA run) there is no statistically
significant difference.

\begin{figure}
  \includegraphics[width=\columnwidth]{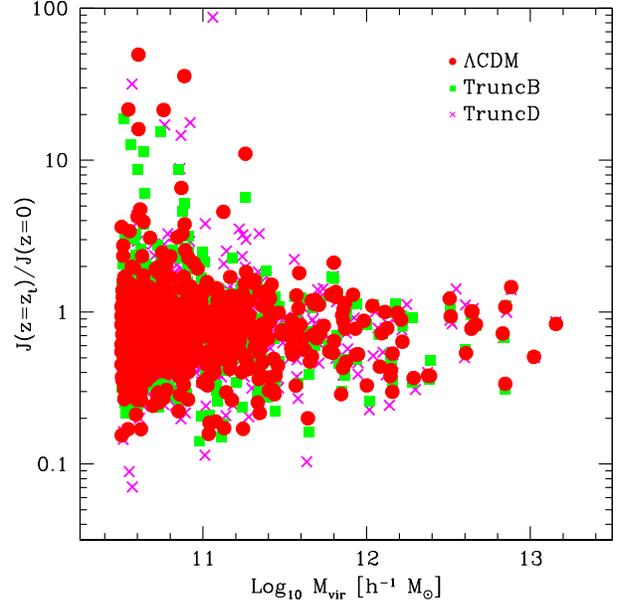}
  \includegraphics[width=\columnwidth]{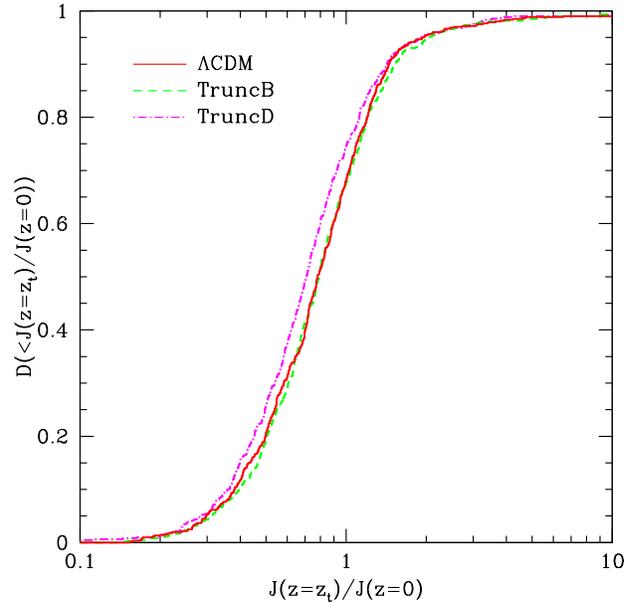}
  \caption{{\bf Angular Momentum at Turnaround.} We track the material
    associated with each halo identified at $z$=0 and compute the
    radial extent and angular momentum of this material as a function
    of redshift in the $\Lambda$CDM, TruncB and TruncD runs. When the material 
    has reached its maximum radial extent, we denote the epoch at which 
    this occurs as turnaround and look at the ratio of the magnitude of 
    angular momentum of the material at this redshift $z_{\rm t}$, 
    $J(z_{\rm t})$, with respect to the magnitude of the angular
    momentum of this material at $z$=0. In the upper panel we show the 
    variation of this ratio with halo mass at $z$=0; in the lower
    panel, we show the cumulative distribution $D(<J(z_{\rm t})/J(z=0))$.}
\label{fig:jturnaround}
\end{figure}

\subsubsection*{Angular Momentum of the Lagrangian Volume} 

 In Figure~\ref{fig:jturnaround} we investigate the angular momentum of
the Lagrangian region corresponding to the virialised halo at $z$=0 and
determine how it evolves with time for haloes with masses in excess of 
$5 \times 10^{10} h^{-1} \rm M_{\odot}$ at $z$=0. In other words, we
track the angular momentum of all the material that contributes to the final
halo at $z$=0. We identify particles at $z$=9 that reside 
within the virial radius at $z$=0 and compute their angular momentum 
$\vec{J}$ using their centre of mass and centre of mass velocity. In 
addition we estimate the mean radial velocity of this material with respect to
the centre of mass velocity and determine the redshift at which it changes
sign from positive to negative (i.e. from expansion to contraction); this
defines the redshift of turnaround $z_t$. This is typically 
between $0.6 \lesssim z \lesssim 4$ for the haloes we consider. This is
equivalent to one of the two empirical measures of turnaround employed by
\citet{sugerman00}.

We expect tidal torques arising from gravitational interaction with the 
surrounding matter distribution to drive the growth of angular momentum
at early times (prior to turnaround) and so it should not be
particularly sensitive to a small scale cut-off in the power spectrum. 
Linear perturbation theory should hold, and the angular momentum of the 
material should grow in proportion to $(1+z)^{-3/2}$ \citep[cf.][]{white84}. 
Therefore, we expect the angular momentum at turnaround to be close to
its maximum value\footnote{\citet{sugerman00} have shown that the
  angular momentum continues to grow `quasi-linearly' after
  turnaround until first shell crossing, at which point it reaches its 
  maximum value.} and the frequency distribution of angular momenta
should be similar in each of the models we have looked at. Linear
perturbation theory no longer provides a good description of angular
momentum growth subsequent to turnaround and non-linear processes
(i.e. mergers) are believed to become more important drivers of angular
momentum evolution during this phase. Therefore, if there are
differences between the models, we would expect them to be apparent in the
ratio of the `peak' angular momentum at turnaround to the final
angular momentum at $z$=0.

In the upper panel of Figure~\ref{fig:jturnaround} we show the
distribution of $J(z_{\rm t})/J(z=0)$ versus halo mass, while in the lower 
panel we show the cumulative distribution $D(<J(z_{\rm t})/J(z=0))$ for
all haloes with masses in excess of $5 \times 10^{10} h^{-1} \rm
M_{\odot}$ at $z$=0. For clarity we consider only the
$\lambda$CDM (filled circles, solid curves), TruncB (filled squares,
dashed curves) and TruncD (crosses, dotted-dashed curves) runs.
The upper panel reveals that, on average, the ratio of $J(z_{\rm
  t})/J(z=0)$ does not vary appreciably with mass and that it is
slightly less than unity (approximately 0.8). In other words, the
magnitude of the total angular momentum of the material at turnaround
is on average smaller than at $z$=0. 

These Figures reveal that the small differences that we observe 
in the spin distributions are also present in the specific angular momentum. 
The median $J(z_{\rm t})/J(z=0)$ differs by $\sim 10\%$ between the 
$\lambda$CDM model and the TruncD run.\\

\section{Summary \& Discussion}
\label{sec:conclusions}

The focus of this paper has been to determine the extent to which suppressing 
the formation of small-scale structure -- low-mass dark matter haloes -- 
affects observationally accessible properties of galaxy-mass 
dark matter haloes. Using cosmological $N$-body simulations, we have 
investigated the spatial clustering of low-mass haloes around galaxy-mass 
haloes, the rate at which these haloes assemble their mass and at which 
they experience mergers, and their angular momentum content in a fiducial 
$\Lambda$CDM model and in truncated ($\Lambda$WDM-like) models. The main 
results of our study can be summarised as follows;

\paragraph*{Large Scale Structure:} Visual inspection of the density 
distribution reveals that the structure that forms in truncated
models is indistinguishable from that in the $\Lambda$CDM model 
on large scales but differs on small scales. Precisely how small this 
scale is depends on $M_{\rm cut}$, the mass scale below which low-mass 
halo formation is suppressed, which we varied between 
$5 \times 10^9 h^{-1} {\rm M_{\odot}}$ and $10^{11} h^{-1} \rm M_{\odot}$.
For $M_{\rm cut}=5 \times 10^9 h^{-1} {\rm M_{\odot}}$ the differences with 
respect to the $\Lambda$CDM model are negligible, but they become significant 
for $M_{\rm cut}=10^{11} h^{-1} \rm M_{\odot}$.

\paragraph*{Spatial Clustering:} These visual differences are apparent in the
clustering strength of lower-mass secondary haloes around galaxy-mass primaries.
Fixing the primary mass at $M_{\rm vir}=10^{11} h^{-1} \rm M_{\odot}$, we 
found that the clustering strength of secondaries around primaries 
depends strongly on $M_{\rm cut}$ and the minimum secondary mass. If we 
include secondaries with masses $M_{\rm vir} \geq 3 \times 10^9 h^{-1} 
\rm M_{\odot}$, the differences are as great as $\sim 50\%$ when $M_{\rm 
cut}$=$10^{11} h^{-1}\, \rm M_{\odot}$. Unsurprisingly, we found no 
dependence on $M_{\rm cut}$ if secondaries are restricted to haloes with 
masses $M_{\rm vir} \geq 10^{11} h^{-1} \rm M_{\odot}$.

\paragraph*{Mass Accretion and Merger Rates:} The sensitivity of the
clustering strength to $M_{\rm cut}$ has immediate consequences for the 
frequency of minor mergers. The effect is most striking for models with 
$M_{\rm cut} \geq 5 \times 10^{10} h^{-1}\,\rm M_{\odot}$, when the rate 
of all mergers with mass ratios in excess of $\sim$$6\%$ is suppressed 
across all redshifts by factors of $\sim$$2$ to $3$ in haloes with 
virial masses of $M_{\rm vir} \lesssim 5 \times 10^{11} h^{-1} \rm 
M_{\odot}$. This effect must be driven by a reduction in the number of 
minor mergers because the frequency of major mergers does not depend on 
$M_{\rm cut}$ other than in haloes with masses $M_{\rm vir} \sim M_{\rm 
cut}$. Interestingly we found that the total mass accretion rate does 
not appear to be sensitive to $M_{\rm cut}$ at all.

\paragraph*{Halo Angular Momentum:} Minor mergers appear to have little 
influence on the angular momentum content of galaxy-mass haloes.

\begin{enumerate}

\item We computed the spin parameter $\lambda$ and found no obvious dependence 
on $M_{\rm cut}$ but a strong dependence on mass accretion history has --
a marked systematic offset is evident between the average spins of haloes 
with violent mass accretion histories and those with quiescent histories -- 
by a factor of $\sim\!2$ to $3$, independent of $M_{\rm cut}$. The spin of 
individual haloes evolve in an almost stochastic fashion over time and on 
average do not show any obvious evolution with redshift.

\item We examined the angular momentum distribution within haloes by
constructing specific angular momentum profiles, which quantify the fraction 
of material within a halo that has specific angular momentum of $j$ or less. 
We found a weak trend for halo material in truncated models with values of
$M_{\rm cut}$ greater than $10^{10} h^{-1}\,\rm M_{\odot}$ to have on average 
smaller values of $j$ by $25\%$ at most, but the r.m.s scatter is large for 
a given $M(<j)$ in all our models and the differences have a low statistical 
significance.

\item We investigated the angular momentum of the Lagrangian region 
corresponding to the virialised halo at $z$=0 and determined how it 
evolves with time. We calculated $J(z_{\rm t})/J(z=0)$, the ratio of the 
angular momentum of the material at the turnaround redshift $z_{\rm t}$ 
to $z$=0. Again the differences between the models are small, at most $10\%$.

\end{enumerate}
\noindent These results indicate that small-scale structure has little 
impact on the angular momentum content of galaxy-mass haloes, in
broad agreement with those of \citet{wang08}, who studied halo formation in
Hot Dark Matter models, and \citet{bullock02} and \citet{chen02}, who 
looked at WDM models.\\

These results show that there are differences in the spatial clustering
and merger rates of low-mass haloes between our fiducial 
$\Lambda$CDM model and the truncated models -- but that they are evident
only in the most extreme truncated models, with $M_{\rm cut}$ in excess of 
$10^{10}h^{-1}{\rm M}_{\odot}$. As we noted in the introduction, this is 
inconsistent with astrophysical constraints on the
putative WDM particle mass. Therefore, measuring the effect on 
spatial clustering or the merger rate is likely to be observationally 
difficult for realistic values of $M_{\rm cut}$, equivalent to our 
TruncA runs, and so isolating the effect of this small-scale structure 
would appear to be remarkably difficult to detect, at least in the present day 
Universe. 

However, there are important caveats. The effect may not be so subtle in the 
high redshift Universe,
during the earliest epoch of galaxy formation, and so we might expect marked
differences in the abundances of low-mass satellites between our fiducial 
$\Lambda$CDM model and WDM or truncated models. This may have observable 
consequences for the ages and metallicities of the oldest stars in galaxies 
\citep[e.g.][]{frebel05}, the abundance of metal poor globular clusters
and the assembly of galaxy bulges and stellar haloes. In addition, there is no compelling
reason to expect that the efficiency of galaxy formation will differ between
a $\Lambda$CDM model and a WDM or truncated model, and so it may be the case
that galaxy formation in a WDM(-like) model is easier to reconcile with the 
observed galaxy population than galaxy formation in the fiducial $\Lambda$CDM
model \citep[see also][]{menci.etal.2012,benson.2013}. We shall return to these
ideas in forthcoming work.

\section*{Acknowledgments}
CP thanks the anonymous referee for their thoughtful report.
This work was supported by ARC DP130100117 and by computational resources on 
the EPIC supercomputer at iVEC through the National Computational Merit 
Allocation Scheme. The research presented in this paper was undertaken as part 
of the Survey Simulation Pipeline 
(SSimPL; {\texttt{http://ssimpl-universe.tk/}).

\bibliographystyle{apj}
\bibliography{paper}

\end{document}